\documentclass[10pt]{article}
\usepackage[dvips]{graphicx}

\usepackage{epsfig}
\usepackage{graphicx}
\usepackage{indentfirst}
\usepackage{amsmath}
\usepackage{amsfonts}
\usepackage{amssymb}

\begin{document}
\begin{center}
{\Large\bf   Exploring Cylindrical Solutions in Modified $f(G)$ Gravity\\}
\medskip

M. J. S. Houndjo$^{(a,b)}$\footnote{e-mail:
sthoundjo@yahoo.fr},\, M. E. Rodrigues$^{(c)}$\footnote{e-mail: esialg@gmail.com},\, D. Momeni\footnote{e-mail: d.momeni@yahoo.com}$^{(d)}$\, and R. Myrzakulov$^{(d)}$\footnote{e-mail:rmyrzakulov@csufresno.edu}\\
$^{a}${ \it Departamento de Engenharia e Ci\^{e}ncias Naturais- CEUNES -
Universidade Federal do Esp\'irito Santo\\
CEP 29933-415 - S\~ao Mateus - ES, Brazil}\\
 $^{b}${\it Institut de Math\'{e}matiques et de Sciences Physiques (IMSP) - 01 BP 613 Porto-Novo, B\'{e}nin}\\
$^{c}$\,{\it Universidade Federal do Esp\'irito Santo -
Centro de Ci\^{e}ncias Exatas - Departamento de F\'isica\\
Av. Fernando Ferrari s/n - Campus de Goiabeiras
CEP29075-910 - Vit\'oria/ES, Brazil}\\
$^{d}${\it
Eurasian International Center for Theoretical Physics -
 Eurasian National University,\\ Astana 010008, Kazakhstan}\\
\date{}

\end{center}
\begin{abstract}
We present cylindrically symmetric solutions for a type of the Gauss-Bonnet gravity, in details. We derive the full system of the field equations and show that there exist seven families of exact solutions for three forms of viable models. By applying the method based on the effective fluid energy momentum tensor components, we evaluate the mass per unit length for the solutions. From dynamical point of the view, by evaluating the null energy condition for these configurations, we show that in some cases the azimuthal pressure breaks the energy condition. This violation of the null energy condition predicts the existence of a cylindrical wormhole.
\end{abstract}

Pacs numbers: 04.50.kd, 04.20.-q
\section{Introduction}
Gravity is the most popular fundamental force of the nature. We know many things about the attractive gravitational interaction from 400 years ago. In the weak field, we can describe the long distance interactions of the solar bodies by classical Newtonian mechanics. But if you want to explain some non classical facts of gravity , we need a relativistic , general covariance formulation of gravity in terms of the classical gauge fields \cite{moshe}. According to the equivalence principle, based on the Mach's idea, the existence of any kind of the matter fields (classical or quantum form) needs a curved space to ignoring the local effects of the gravity in any local Lorentz invariance frame of the coordinates. The simplest one is based on the Ricci scalar $R$ curvature , a second order derivatives of the metric of the spacetime manifold, adopted by the uniqueness theorem. It works in the best conditions in solar system, but it needs some necessary modifications beyond the galaxy and also inside the disk's galaxy to explain the problems like dark matter or dark energy. A class of all such alternative gravitational models is called modified gravity (see \cite{sergei1} for an excellent review). Different kinds of this modifications has been proposed. The first extension which remains still now, as a remarkable one is obtained just by replacing the curvature scalar $R$ by an arbitrary function $f(R)$ \cite{buchdahl}. The simple case of $R^2$ can be used in the inflationary scenario \cite{Starobinsky}, and also to explain the effects of the observational constraints on solar system . Also, a wide class of the cosmological aspects have been investigated recently based on such modifications \cite{fr1}. But, the modifications of the Einstein gravity is not restricted to the curvature corrections. Gauss-Bonnet topological invariant $G$ gives us another possibility, which has a rich class of the cosmological predictions \cite{sergei2}. Also, the black objects in a theory of gravity with both $R,G$ in the form of $f(R,G)$ have been obtained in the literatures \cite{gbbh}.\\
 The cylindrical symmetry gives us another opportunity to obtain exact solutions. In Einstein gravity it exists a wide class of exact solutions with cylindrical symmetry in vacuum and also with cosmological constant and scalar fields \cite{einstein-cylind}. In modified gravity different kinds of exact solutions with cylindrical symmetry have been discussed in the literature, in f(R) \cite{cylinder-f(R)}, Horava-Lifshitz gravity \cite{cylinder-hl}, f(T) \cite{cylinder-ft} and etc.
 Recently \cite{cylinder-gb1} we start obtaining exact solutions in cylindrical symmetric spacetimes but in the context of $f(R,G)$ gravity . We derived the field equations for a general form of the modified f(G) models with Lagrangian in the form of $f(R,G)\equiv R+f(G)$. In \cite{cylinder-gb1}, we proved that there exists a non vacuum family of exact solutions with $R=0$ but with $G\neq0$, corresponding to a fluid with generic forms of the energy density $\rho$ and the pressure components $p_r,p_{\varphi},p_z$. The solution  resembles the case of the vacuum Levi-Civita (LC) family \cite{LC} in the Einstein gravity as the exterior metric of a cosmic string \cite{vilenkin,linet,tian}. In that letter , we just expose a very specific solution which corresponds to the LC solution. In this present paper we continue studying the exact solutions for $f(R,G)\equiv R+f(G)$, finding more solutions for some viable forms of $f(G)$, by solving the differential equations of the motion analytically. Moreover, we separately discussed the solutions for three viable forms of $f(G)$. In any case, we discussed the viability and all possible non trivial solutions of the system which is under review. Also, we examine the solutions from the energy conditions using the modified (effective) gravitational field equations. As a review, we point out here the definitions of the null energy condition
(NEC), weak energy condition (WEC), strong energy condition (SEC) and the dominant energy
condition (DEC) as follows \cite{lobo,anzhong}
\begin{eqnarray}
\text{NEC}&\Longleftrightarrow&\rho_{\text{eff}}+p_{\text{eff}}\geq0.\label{n1}\\
\text{WEC}&\Longleftrightarrow& \rho_{\text{eff}}\geq0\ \text{and}\ \rho_{\text{eff}}+p_{\text{eff}}\geq0.\label{n2}\\
\text{SEC}&\Longleftrightarrow& \rho_{\text{eff}}+3p_{\text{eff}}\geq0\ \text{and}\ \rho_{\text{eff}}+p_{\text{eff}}\geq0.\label{n3}\\
\text{DEC}&\Longleftrightarrow& \rho_{\text{eff}}\geq0\ \text{and}\ \rho_{\text{eff}}\pm p_{\text{eff}}\geq0.\label{n4}
\end{eqnarray}
By starting from any gravitational model and by rewriting it in the forms of a set of effective energy momentum components, we will find that the essence of these energy conditions is independent from the form of the action \cite{hawking}. It is easy to find that always, NEC implies WEC and WEC implies SEC and DEC. In all the subsequent models we will assume that the regular matter  satisfies all the energy conditions separately i.e. $\rho\geq0$, $\rho\pm p_i\geq0$, $\rho+3p_i\geq0$. In literature, for example in $f(R)$ gravity it has been tested for validity of the energy conditions \cite{wang1}. Also, in $f(T)$ we examined these conditions for some viable models \cite{ecft}. Thus, we must check the validity of these conditions  for our cylindrical $f(G)$ solutions. For the importance of the NEC , we will check just this energy condition in any case. Further we obtained the formula for mass per unit length  by the method proposed by Israel \cite{israel}. The Israel's formula gives us the mass per length for any isolated cylindrically symmetric system filled by the energy momentum of the matter as the following formula:
\begin{eqnarray}
\text{Mass per unit length }&\Longleftrightarrow&m=2\pi\int{\sqrt{-g}dr(\rho+\Sigma p_{i})}.\label{m}
\end{eqnarray}
Recently this formula has been used to calculate the quasi-local mass per unit length for a class of the conformally cylindrical metrics as the interior solutions to the Linet-Tian family \cite{conf}. Our plan in this paper is the following: In section II, we present the model and we derive field equations and energy conditions inequalities. In sections III, IV, V we will solve the system of equations for some particular viable models of $f(G)$. In section VI, we analysis the dynamics of the solutions using energy conditions. We conclude and summarize in final section our results.

\section{Formalism of $f(G)$ gravity and equations of motion within cylindrical metric}
We introduce the action for a typical general $R+f(G)$ modification  of gravity following  \cite{sergei3}
\begin{eqnarray}
S=\int d^4x\sqrt{-g}\left[\frac{R}{2\kappa}+f(G)\right]+S_m\,\,,\label{ratbay1}
\end{eqnarray}
As usual in Einstein gravity, by working with the commutative connections in a Riemannian spacetime, $R$ represents the Ricci scalar, and the modification function $f(G)$ corresponds to  a generic globally differentialable function of the Gauss-Bonnet topological invariant $G$. Also we add the matter action   $S_m$ which induces  the energy momentum tensor $T_{\mu\nu}$ . For convention we take  $\kappa=8\pi \mathcal{G}_N$, where $\mathcal{G}_N$ is nothing just the usual classical Newtonian gravitational coupling. In metric formalism and by taking the metric as the dynamical variable of the model,   the equations of motion for
the metric $g_{\mu\nu}$ from (\ref{ratbay1}) read 
\begin{eqnarray}
R_{\mu\nu}-\frac{1}{2}Rg_{\mu\nu}+8\Big[R_{\mu\rho\nu\sigma}+R_{\rho\nu}g_{\sigma\mu}
-R_{\rho\sigma}g_{\nu\mu}-R_{\mu\nu}g_{\sigma\rho}+R_{\mu\sigma}g_{\nu\rho}\nonumber\\
+\frac{R}{2}\left(g_{\mu\nu}g_{\sigma\rho}-g_{\mu\sigma}g_{\nu\rho}\right)\Big]\nabla^{\rho}\nabla^{\sigma}f_{G}+\left(Gf_{G}-f\right)g_{\mu\nu}=\kappa T_{\mu\nu}\,\,,\label{ratbay2}
\end{eqnarray}
We denote by $f_{G}=\frac{df(G)}{dG}$ also the Gauss-Bonnet (GB) term is defined as $G=R^2-R_{\mu\nu}R^{\mu\nu}+R_{\mu\nu\lambda\sigma}R^{\mu\nu\lambda\sigma}$, $R_{\mu\nu}$ and $R_{\mu\nu\lambda\sigma}$ play the roles of  the Ricci tensor and Riemann tensors, respectively. WE adopted the signature of the Riemannian metric as $(+---)$ , and $\nabla_{\mu}V_{\nu}=\partial_{\mu}V_{\nu}-\Gamma_{\mu\nu}^{\lambda}V_{\lambda}$ and $R^{\sigma}_{\;\mu\nu\rho}=\partial_{\nu}\Gamma^{\sigma}_{\mu\rho}-\partial_{\rho}
\Gamma^{\sigma}_{\mu\nu}+\Gamma^{\omega}_{\mu\rho}\Gamma^{\sigma}_{\omega\nu}
-\Gamma^{\omega}_{\mu\nu}\Gamma^{\sigma}_{\omega\rho}$ for the covariant (curved)derivative and the Riemann tensor, respectively. We have different possibilities for the form of a static cylindrically symmetric metric. The more common and simplest on is the Weyl gauge in which it is easy to interpret the metric functions as the generalized potential functions. This form of the coordinates, called Weyl coordinates $(t,r,\varphi,z)$ is given by \cite{Stephani}
\begin{eqnarray}
g_{\mu\nu}=diag\left\{e^{2u(r)},-e^{2k(r)-2u(r)}, -w(r)^2e^{-2u(r)},-e^{2k(r)-2u(r)}\right\}\,\,.\label{ratbay3}
\end{eqnarray}
Here we must be very careful about any simple interpretation of the coordinates $r,t$ as the usual radial and time. The curvature scalar and the Gauss-Bonnet invariant term $G$ read, respectively
\begin{eqnarray}
R=2\left(\frac{w''}{w}-\frac{u'w'}{w}-u''+u'^2+k'' \right)e^{2u-2k}\,\,,\label{ratbay6}\\
G=\frac{8e^{4(u-k)}}{w}\Big[ k'u'w''-u'^2w''-2u'u''w'+k'u''w'-u'^3w'+3k'u'^2w'+k''u'w'-2k'^2u'w'\nonumber\\+w\left(3u'^2u''-2k'u'u''+2u'^4-4k'u'^3-k''u'^2+2k'^2u'^2\right)\Big]\label{ratbay7}
\end{eqnarray}
So by applying the static cylindrical metric (\ref{ratbay3}) the equation (\ref{ratbay2}), we obtain the following system of coupled four non linear  differential equations
\begin{eqnarray}
\frac{8e^{4(u-k)}}{w}\Big\{ \left[u'w'-k'w'-w\left(u'^2-k'u'\right)\right]f''_{G}+
\Big[u'w''-k'w''+u''w'-2k'u'w'-k''w'\nonumber\\
+2k'^2w'-w\left(2u'u''-k'u''+u'^3-3k'u'^2-k''u'+2k'^2u'\right)\Big]f'_{G}
-\Big[u'^2w''-k'u'w''+2u'u''w'\nonumber\\
-k'u''w'+u'^3w'-3k'u'^2w'-k''u'w'+2k'^2u'w'-w\Big(3u'^2u''-2k'u'u''+2u'^4-4k'u'^3-k''u'^2\nonumber\\
+2k'^2u'^2\Big)\Big]f_{G}\Big\}+\frac{e^{2(u-k)}}{w}\Big[2u'w'-w''+w\left(2u''-u'^2-k''
\right)\Big]-f=\kappa\rho\,\,,\label{ratbay8}\\
\frac{8e^{4(u-k)}}{w}\Big\{ 3\left[u'^2w'+k'u'w'+w\left(u'^3-k'u'^2\right)\right]f'_G
+\Big[u'^2w''-k'u'w''+2u'u''w'-ku''w'+u'^3w'\nonumber\\
-3k'u'^2w'-k''u'w'+2k'^2u'w'+w\Big(2k'u'u''-3u'^2u''-2u'^4+4k'u'^3+k''u'^2-k'^2u'^2\Big)
\Big]f_{G}\Big\}\nonumber\\
+\frac{e^{2(u-k)}}{w}\Big[k'w'-wu'^2\Big]+f=\kappa p_{r}\,\,,\label{ratbay9}
\end{eqnarray}
\begin{eqnarray}
\frac{8e^{4(u-k)}}{w}\Big\{w\left(k'u'-u'^2\right)f''_{G}+w\Big[k'u''-2u'u''-3u'^3+5k'u'^2+k''u'-2k'^2u'\Big]f'_{G}\nonumber\\
+\Big[u'^2w''-k'u'w''+2u'u''w'-k'u''w'+u'^3w'-3k'u'^2w'-k''u'w'+2k'^2u'w'
+w\Big(2k'u'u''\nonumber\\-3u'^2u''-2u'^4+4k'u'^3+k''u'^2-2k'^2u'^2\Big)\Big]f_{G}\Big\}
+e^{2(u-k)}\left(u'^2+k''\right)+f=\kappa p_{\varphi}\,\,,\label{ratbay10}\\
\frac{8e^{4(u-k)}}{w}\Big\{ \left(u'w'-wu'^2\right)f''_G+\Big[u'w''+u''w'+2u'^2w-3k'u'w'
+w\Big(3k'u'^2-2u'u''-3u'^3\Big)\Big]f'_G\nonumber\\
+\Big[u'^2w''-k'u'w''+2u'u''w'-k'u''w'+u'^3w'-3k'u'^2w'-k''u'w'+2k'^2u'w'+w\Big(2k'u'u''\nonumber\\
-3u'^2u''-2u'^4+4k'u'^3+k''u'^2-2k'^2u'^2\Big)\Big]f_G\Big\}+\frac{e^{2(u-k)}}{w}\left(w''-k'w'+wu'^2\right)+f=\kappa p_z\,\,.\label{ratbay11}
\end{eqnarray}
Using the Eqs.(\ref{ratbay8}-\ref{ratbay11}) we have the following expression for mass per unit length, through the (\ref{m}),
\begin{eqnarray}
m&=&\frac{\pi}{2\kappa^2}\int_0^{r_0}{dr  e^{-2 (k+3u)}\Big(4 w e^{4 (k+u)}(2 f+w (f_{G}''u'
   (k'-u')+f_{G}' u''(k'-2u'))+u' (k''+5 k'
   u'-2 k'^2-3 u'^2))\Big)}\nonumber\\&&-\frac{\pi}{2\kappa^2}\int_0^{r_0}{dr \Big(32 e^{8 u}
   (f_{G}'' w' (k'-2 u')+f_{G}'
   (w'' (k'-2 u')+w' (k''+2 k'
   u'-2 k'^2-2 u''-3 u'^2))}\nonumber\\&&+\frac{\pi}{2\kappa^2}\int_0^{r_0}{dr w
   (f_{G}'' u' (2 u'-k')+f_{G}'
   (u' (-(3 k'+2) u'+k' (2
   k'-1)+u'^2)-u'' (k'-4u'))))}\nonumber\\&&+\frac{\pi}{2\kappa^2}\int_0^{r_0}{dre^{8 k} (32-f_{G}Gw^2)+8e^{2 k+6 u}(wu''+u'w')\Big)}
\end{eqnarray}
Here $r_0$ denotes the radius of a typical string. We can read this integral without the upper bound $r_0$ as a well defined expression for local mass (quasi local) concentrated in a shell with radius $r$ of the axis. Also, the singularity of the integral as an improper integral at $r=0$ understood on the meaning of the thick and thin strings. Also, we used the expressions of the energy-momentum tensor components effective in the integral. This definition is conventional and it can be understood easily if we write the equations in the weak field of a Newtonian distribution of the matter and usage of the classical Poisson's equation for a non localized distribution of the matter with $T_{\mu\nu}$ in a finite (or compact, spatially) region $\Omega$. So, the integral must be evaluated as integral over spacelike hypersurface orthogonal sheet in $\Omega$ as $\int_{\Omega}{ dr\sqrt{h}}$ where $h$ denotes the determinant of the positive definite space like slice of the metric manifold $\Sigma_4$.

Also, the NEC for three different pressure components implies that we must check the following inequalities:
\par
NEC for radial pressure $p_r$ \par

\begin{eqnarray}
&&f_GG-\frac{1}{8} f_GGwe^{4(k-u)}-\frac{e^{2(u-k)}}{w}(wk''-k'w'-2 wu''-2 u'
   w'+2 wu'^2+w'')-\nonumber\\&&\frac{8 e^{4(u-k)}}{w}\Big[-w f_G'' k' u'+f_G'' k'
   w'-f_G'' u' w'+wf_G''
   u'^2+f_G' k'' w'-wf_G' k'
   u''-f_G' k' u' w'\nonumber\\&&+2 wf_G' k'^2
   u'-wf_G' k' u'+f_G' k' w''-2
   f_G' k'^2 w'-f_G' u'' w'-f_G'
   u' w''-3 f_G' u'^2 w'\nonumber\\&&-2 wf_G'
   u'^3+2 w f_G' u' u''\Big]\geq0 \label{NEC1}
\end{eqnarray}

\par
NEC for azimuthal pressure $p_\varphi$\par

\begin{eqnarray}
&&\frac{e^{-4(k+u)}}{8w}\Big[we^{4 (k+u)} (w (f_G''u'(k'-u')+f_G' u''(k'-2u'))\nonumber\\&&+f_G G+u'(k''+5 k' u'-2 k'^2-3u'^2))\nonumber\\&&-64 e^{8u} (f_G'' w'(k'-u')+f_G' (w''(k'-u')+w' (k''+2 k'(u'-k')-u''))\nonumber\\&&+w
   (f_G'' u' (u'-k')+f_G'(2k'^2 u'-k'(u''+3u'^2+u')+u'^3+2 u'
   u'')))\nonumber\\&&+e^{8 k} (64-f_G Gw^2)+8 e^{2 k+6 u} (2 w u''+2 u'w'-w'')\Big]\geq0\label{NEC2}
\end{eqnarray}

\par
NEC for axial pressure $p_z$\par

\begin{eqnarray}
&&\frac{e^{-2(k-u)}}{w}(wk''+k'w'-2 wu''-2 u' w')-\frac{8e^{4(u-k)}}{w}\Big[-w f_G'' k' u'+f_G'' k' w'\nonumber\\&&-2
   f_G'' u' w'+2 w f_G''
   u'^2+f_G' k'' w'\nonumber\\&&-w f_G' k' u''+5
   f_G' k' u' w'-6 w f_G' k' u'^2+2
   w f_G' k'^2 u'-w f_G' k'
   u'+f_G' k' w''\nonumber\\&&-2 f_G' k'^2 w'-2
   f_G' u'' w'-2 f_G' u' w''+4 w
   f_G' u'^3-2 w f_G' u'^2+4 w
   f_G' u' u''\Big]\geq0\label{NEC3}
\end{eqnarray}
We will check NEC through (\ref{NEC1}-\ref{NEC3}) for the exact solutions of the viable models.

\section{Solutions for the model $f(G)=\alpha G^2$}
This is a special case of the models which have been discussed before \cite{cognola}. This model is interesting because the Big-Rip singularity may not appear. Also, the cosmology of this model predicts the existence of a transient phantom epoch, compatible with the observational data. Moreover, this viable model is the dominant term of a more general case of $f(G)=c_1G^{\beta_1}+c_2G^{\beta_2}$, for $\beta_1>\beta_2$ in the regime of the high curvature and also in the primordial formation structure era. In this case the system of the equations of motion (\ref{ratbay8}-\ref{ratbay11}) are integrable in five different classes as follows:
Here we propose solutions to this model, i.e, finding solutions to the line element and expressions to the energy density and the pressures. We present the solutions to the line element as

\par
$\bullet$ First case\par
\begin{eqnarray}
u(r)=\ln{\left[w(r)+u_1r\right]}\,,\quad k(r)=\ln{\left(w(r)\right)}\,,\quad w(r)=w_0+w_1r\,\,,
\end{eqnarray}
where $u_1$, $w_0$ and $w_1$ are  constants. Then, the Gauss-Bonnet invariant and the curvature scalar become respectively
\begin{eqnarray}
G(r)&=&-\frac{1}{(w_0+w_1r)^7}\Big\{8u_1^2w_0^2\left(u_1+w_1\right)\left[4w_1(w_0+w_1r)+u_1(w_0+4w_1r)
\right]\Big\}\\
R(r)&=&\frac{1}{(w_0+w_1r)^4}\Big\{2u_1w_0\left[3w_1(w_0+w_1r)+u_1(2w_0+3w_1r)\right]\Big\}
\end{eqnarray}
By making use of the field equations (\ref{ratbay8})-(\ref{ratbay11}), one gets the energy density and pressures as
\begin{eqnarray}
\rho(r)=-\frac{u_1w_0}{\kappa^2(w_0+w_1r)^{15}}\Big\{\left(w_0+w_1r\right)^{11}\left[4w_1\left(w_0+w_1r\right)+
u_1\left(3w_0+4w_1r\right)\right]\nonumber\\
-64\alpha u_1^3w_0^3\left(u_1+w_1\right)\Big[544w_1^3\left(w_0+w_1r\right)^3+
96u_1w_1^2\left(w_0+w_1r\right)^2\left(w_0+17w_1r\right)\nonumber\\
+u_1^3\left(w_0^3+15w_0^2w_1r+96w_0w_1^2r^2+544w_1^3r^3\right)+3u_1^2w_1\left(5w_0^3
+69w_0^2w_1r+608w_0w_1^2r^2+544w_1^3r^3\right)\Big]\Big\}
\end{eqnarray}
\begin{eqnarray}
p_r(r)=\frac{u_1w_0}{\kappa^2\left(w_0+w_1r\right)^{14}}\Big\{-\left(w_0+w_1r\right)^{10}\left[
2w_1\left(w_0+w_1r\right)+u_1\left(w_0+2w_1r\right)\right]\nonumber\\
+64\alpha u_1w_0\left(u_1+w_1\right)^2\Big[288w_1^4\left(w_0+w_1r\right)^4+36u_1w_1^3
\left(w_0+w_1r\right)^3\left(9w_0+32w_1r\right)\nonumber\\
+4u_1^2w_1^2\left(w_0+w_1r\right)^2\left(41w_0^2+243w_0w_1r+432w_1^2r^2\right)\nonumber\\
+u_1^4\left(-w_0^4+10w_0^3w_1r+164w_0^2w_1^2r^2+324w_0w_1^3r^3+288w_1^4r^4\right)\nonumber\\
2u_1^2w_1\left(5w_0^4+169w_0^3w_1r+650w_0^2w_1^2r^2+1062w_0w_1^3r^3+576w_1^4r^4
\right)\Big]\Big\}
\end{eqnarray}
\begin{eqnarray}
p_{\varphi}(r)=-\frac{u_1w_0}{\kappa^2\left(w_0+w_1r\right)^{14}}\Big\{-\left(w_0+w_1r\right)^{10}\left[2w_1
\left(w_0+w_1r\right)+u_1\left(w_0+2w_1r\right)\right]\nonumber\\
-64\alpha u_1^2w_0^2\left(u_1+w_1\right)^2\Big[432w_1^3\left(w_0+w_1r\right)^3+
4u_1w_1^2\left(w_0+w_1r\right)^2\left(-13w_0+324w_1r\right)\nonumber\\
-u_1^3\left(w_0^3+14w_0^2w_1r+52w_0w_1^2r^2-432w_1^3r^3\right)\nonumber\\
+2u_1^2w_1\left(-7w_0^3-59w_0^2w_1r+596w_0w_1^2r^2+648w_1^3r^3\right)\Big]\Big\}
\end{eqnarray}
\begin{eqnarray}
p_z(r)=-\frac{u_1w_0}{\kappa^2\left(w_0+w_1r\right)^{14}}\Big\{-\left(w_0+w_1r\right)^{10}
\left[2w_1\left(w_0+w_1r\right)+u_1\left(w_0+2w_1r\right)\right]\nonumber\\
-64\alpha u_1w_0\left(u_1+w_1\right)^2\Big[96w_1^3\left[w_0+(r-1)w_1\right]
\left(w_0+w_1r\right)^4\nonumber\\
+12u_1w_1^2\left(w_0+w_1r\right)^3\left[9w_0^3+w_0w_1\left(27+41r\right)
+32w_1^2r(r-1)\right]\nonumber\\
+4u_1^2w_1\left(w_0+w_1r\right)^2\left[3w_0^3+4w_0^2w_1\left(21r-4\right)+9w_0w_1^2r\left(
27+25r\right)+144w_1^3r^2(r-1)\right]\nonumber\\
+u_1^4\left[-w_0^4+2w_0^3w_1r(6r-7)+8w_0^2w_1^2r^2(15r-8)+12w_0w_1^3r^3(27+17r)+96w_4r^4(r-1)\right]
\nonumber\\
2u_1^3w_1\Big((12r-7)w_0^4+w_0^3w_1r(186r-71)+2w_0^2w_1^2r^2(211+264r)\nonumber\\
+42w_0w_1^3r^3(7+13r)+192w_1^4r^4(r-1)\Big)\Big]\Big\}
\end{eqnarray}
Now the metric reads:
\begin{equation}
ds^2=\left[w(r)+u_1r\right]^2dt^2-\frac{w(r)^2}{\left[w(r)+u_1r\right]^2}\Big(dr^2+dz^2+d\varphi^2\Big)
\end{equation}
This is a conform-stationary metric, non singular, so cannot describe a black string.\\
The mass expression in this case reads from (\ref{m}), and by integration we obtain
\begin{eqnarray}
m&=&-\frac{\Delta}{60 {w_1} ({r_0}
   {w_1}+{u_1}+{w_0})^5}-\frac{48 {u_1} \log ({r_0}
   {w_1}+{u_1}+{w_0})}{{w_1}}+8
   {r_0}-\nonumber\\&&\frac{{w_1}^2(7 {u_1}^3+28 {u_1}^2
   {w_0}+24 {u_1} {w_0}^2+12 {w_0}^3)}{12
   ({u_1}+{w_0})^4}+\frac{4 {u_1}^2 (87
   {u_1}^4+385 {u_1}^3 {w_0}+650 {u_1}^2
   {w_0}^2+500 {u_1} {w_0}^3+150 {w_0}^4)}{5
   {w_1} ({u_1}+{w_0})^5}\nonumber\\&&+\frac{2 {u_1}
   {w_1}}{{u_1}+{w_0}}+\frac{48 {u_1} \log
   ({u_1}+{w_0})}{{w_1}},\\
   \Delta &=&-720 {u_1}^5 ({r_0} {w_1}+{u_1}+{w_0})+2400
   {u_1}^4 ({r_0} {w_1}+{u_1}+{w_0})^2\nonumber\\&&+45
   {u_1}^3 {w_1}^3 ({r_0}
   {w_1}+{u_1}+{w_0})-4800 {u_1}^3 ({r_0}
   {w_1}+{u_1}+{w_0})^3-80 {u_1}^2 {w_1}^3
   ({r_0} {w_1}+{u_1}+{w_0})^2\nonumber\\&&+7200 {u_1}^2
   ({r_0} {w_1}+{u_1}+{w_0})^4+60 {u_1} {w_1}^3
   ({r_0} {w_1}+{u_1}+{w_0})^3-60 {w_1}^3
   ({r_0} {w_1}+{u_1}+{w_0})^4\nonumber\\&&+120 {u_1}
   {w_1}^2 ({r_0} {w_1}+{u_1}+{w_0})^4+96
   {u_1}^6
\end{eqnarray}
Because the mass of the cosmic string is about $m\sim10^{-7}$ we can constraint the parameters $w_1,u_1,...$ as viable functions of the observational quantities.

\par
$\bullet$ Second case\par
The line element in this case are presented as
\begin{eqnarray}
w(r)=w_0\,,\quad k(r)=\ln{\left[w(r)\right]}\,,\quad u(r)=\ln{\left[w(r)+u_0r^2\right]}\,\,,
\end{eqnarray}
where $u_0$ and $w_0$ is a constant but may not be confused with that of the previous case. With these solutions, the Gauss-Bonnet invariant reads
\begin{eqnarray}
G(r)=\frac{64u_0^3r^2\left(3w_0+u_0r^2\right)}{w_0^4}
\end{eqnarray}
and the energy density and pressures are the following expressions
\begin{eqnarray}
\rho(r)=-\frac{4u_0}{\kappa^2w_0^8}\Big(11264\alpha u_0^7r^8+32768\alpha u_0^6w_0r^6+
35840\alpha u_0^5w_0^2r^4+18432\alpha u_0^4w_0^3r^2+2u_0w_0^6r^2-w_0^7\Big)
\end{eqnarray}
\begin{eqnarray}
p_r(r)=\frac{4u_0^2r^2}{\kappa^2w_0^8}\Big(23552\alpha u_0^6r^6+55296\alpha u_0^5w_0r^4+
27648\alpha u_0^4w_0^2r^2-w_0^6\Big)
\end{eqnarray}
\begin{eqnarray}
p_{\varphi}(r)=-\frac{4u_0^2r^2}{w_0^8\kappa^2}\Big(29696\alpha u_0^6r^6+86016\alpha u_0^5w_0r^4+78848\alpha u_0^4w_0r^2+18432\alpha u_0^3w_0^3-w_0^6\Big)
\end{eqnarray}
\begin{eqnarray}
p_z(r)=\frac{4u_0^2r^2}{\kappa^2w_0^8}\Big(-29696\alpha u_0^6r^6+8192\alpha u_0^6r^7-86016\alpha u_0^5w_0r^4+28672\alpha u_0^5w_0r^5
-78848\alpha u_0^4w_0^2r^2\nonumber\\+32768\alpha u_0^4w_0^2r^3-18432\alpha u_0^3w_0^3+
12288\alpha u_0^3w_0^3r+w_0^6\Big)
\end{eqnarray}
Consequently, the metric reads:
\begin{eqnarray}
ds^2=\left[w_0+u_0r^2\right]^2dt^2-\frac{w_0^2}{\left[w_0+u_0r^2\right]^2}\Big(dr^2+dz^2+d\varphi^2\Big)
\end{eqnarray}
This again represents a conform-stationary metric, non singular, so it can not describe a black string.\\
Like the previous case, mass expression in this case reads from (\ref{m}), and by integration we obtain
\begin{eqnarray}
m&=&\frac{4}{w_0^5}\Big(\frac{r_0^2 u_0^2 w_0^8}{(r_0^2 u_0 + w_0)^4}+\frac{r_0 w_0^6\mathcal{A}_1}{640 (r_0^2 u_0 + w_0)^5}+\mathcal{A}_2+\frac{256\alpha r_0 u_0^4w_0^2\mathcal{A}_3}{15}+\mathcal{A}_4  \Big)
\end{eqnarray}
\begin{eqnarray}
\mathcal{A}_1&=&315 r_0^8 u_0^4 + 1470 r_0^6 u_0^3 w_0 + 2688 r_0^4 u_0^2 w_0^2 + 
 2370 r_0^2 u_0 w_0^3 + 965 w_0^4,\\
 \mathcal{A}_2&=&1/4 u_0 w_0^5 (-1 + w_0^4/(r_0^2 u_0 + w_0)^4 + (4 r_0 w_0)/(r_0^2 u_0 + w_0)) + 
 2048/15 r_0^5 (-21 + 10 r_0) u_0^6 \alpha\nonumber\\&& + 
 1024/3 r_0^3 (-16 + 9 r_0) u_0^5 w_0 \alpha\\
 \mathcal{A}_3&=&-720+\frac{w_0\mathcal{B}}{(r_0^2 u_0 + w_0)^5}\\
 \mathcal{A}_4&=&\frac{{w_0}^{5/2} (63 {w_0}^3-32768 \alpha  {u_0}^4 ({w_0}-32)) \tan
   ^{-1}(\frac{{r_0} \sqrt{{u_0}}}{\sqrt{{w_0}}})}{128 \sqrt{{u_0}}}\\
 \mathcal{B}&=&-15 r_0^8 u_0^4 (-16 + w_0) + 10 r_0^6 u_0^3 (96 - 5 w_0) w_0 + 
 r_0^4 u_0^2 (1440 - 37 w_0) w_0^2 \nonumber\\&&+ 15 w_0^4 (16 + w_0) + 
 5 r_0^2 u_0 w_0^3 (192 + 5 w_0)
\end{eqnarray}

\par
$\bullet$ Third case\par
The line element parameters $w(r)$ and $k(r)$ are fixed as 
\begin{eqnarray}
w(r)=w_0\,\quad k(r)=\ln{\left[w(r)\right]}\,\,.
\end{eqnarray}
Therefore, the Gauss-Bonnet invariant and the curvature become
\begin{eqnarray}
G(r)=\frac{8u'^2(r)}{w_0^4}\left[2u'^2(r)+u''(r)\right]e^{4u(r)},\quad 
R=\frac{2}{w_0^2}\left[u'^2(r)-u''(r)\right]e^{2u(r)}\,\,.
\end{eqnarray}
In order write put down a suitable expression for the line element parameter $u(r)$, we assume the following relation between $G(r)$ and $R(r)$
\begin{eqnarray}
G(r)=e^{2u(u)}u'^2(r)R(r)\,\,.
\end{eqnarray}
by solving this latter, one gets
\begin{eqnarray}
u(r)=u_0+\frac{12+w_0^2}{w_0^2-8}\ln{\left(r\right)}\,\,,
\end{eqnarray}
where $u_0$ is an integration constant, leading to the final forms of $G(r)$ and $R(r)$ as
\begin{eqnarray}
G(r)=-\frac{8(w_0^2-48)(w_0^2+12)^3e^{4u_0}r^{\frac{80}{w_0^2-8}}}{w_0^4(w_0^2-8)^4} ,\label{GB26}
\end{eqnarray}
and 
\begin{eqnarray}
R(r)=\frac{4(w_0^2+12)(w_0^2)e^{2u_0}r^{\frac{40}{w_0^2-8}}}{w_0^2(w_0^2-8)^2}\,\,.
\end{eqnarray}
The energy density and pressures are get respectively as
\begin{eqnarray}
\rho(r)=\frac{(12+w_0^2)e^{2u_0}r^{\frac{40}{w_0^2-8}}}{\kappa^2w_0^8(w_0^2-8)^8}\Big\{-w_0^6(w_0^2-8)^6(3w_0^2-4)\nonumber\\+
64\alpha e^{6u_0}r^{\frac{120}{w_0^2-8}}(w_0^2+12)^4\left(-863232+35072w_0^2-404w_0^4+w_0^6\right)\Big\}
\end{eqnarray}
\begin{eqnarray}
p_r(r)=\frac{(12+w_0^2)^2e^{2u_0}r^{\frac{40}{w_0^2-8}}}{\kappa^2w_0^8(w_0^2-8)^8}\Big\{w_0^6(w_0^2-8)^6\nonumber\\+
64\alpha e^{6u_0}r^{\frac{120}{w_0^2-8}}(w_0^2+12)^4\left(-20736+384w_0^2+w_0^4\right)\Big\}
\end{eqnarray}
\begin{eqnarray}
p_{\varphi}(r)=-\frac{(12+w_0^2)^2e^{2u_0}r^{\frac{40}{w_0^2-8}}}{\kappa^2w_0^8(w_0^2-8)^8}\Big\{-w_0^6(w_0^2-8)^6\nonumber\\+
64\alpha e^{6u_0}r^{\frac{120}{w_0^2-8}}(w_0^2+12)^3\left(1102848-21248w_0^2-84w_0^4+w_0^6\right)\Big\}
\end{eqnarray}
\begin{eqnarray}
p_z(r)&=&-\frac{(12+w_0^2)^2e^{2u_0}r^{\frac{40}{w_0^2-8}}}{\kappa^2w_0^8(w_0^2-8)^8}\Big\{-w_0^6(w_0^2-8)^6
+64\alpha e^{6u_0}r^{\frac{112}{w_0^2-8}}(w_0^2-48)(w_0^2+12)^3\nonumber\\
&\times &\Big[320(w_0^2-8)r^{\frac{w_0^2}{w_0^2-8}}+(w_0^4-36w_0^2-22976)r^{\frac{8}{w_0^2-8}}
\Big]\Big\}
\end{eqnarray} 
Consequently, the metric reads:
\begin{eqnarray}
ds^2=\tilde{r}^{2\beta}d\tau^2-\tilde{r}^{-2\beta}\mu^2\Big(d\tilde{r}^2+d\tilde{z}^2+d\tilde{\varphi}^2\Big),\\ \nonumber \beta\equiv\frac{12+w_0^2}{w_0^2-8},\ \  \tilde{r}=\mu^{\beta^{-1}} r, \tilde{z}=\mu^{\beta^{-1}}z, \tilde{\varphi}=\mu^{\beta^{-1}}\varphi.
\end{eqnarray}
Here $\mu=w_0^{\beta}e^{-\beta u_0}$ denotes the unique parameter of the solution and it may play the role of the conical deflection parameter.
This again represents a conform-stationary metric, non singular, so it can not describe a black string.
Like the previous case, mass expression in this case reads from (\ref{m}), and by integration we obtain
\begin{eqnarray}
m&=&\frac{1}{6r^7}\Big(-\frac{192 \alpha  \beta ^5 (2 \beta -3) (5 \beta  (2 \beta -5)+14) e^{2 {u_0}} r^{2 \beta }}{2 \beta -7}+\frac{48 e^{-6 {u_0}} {w_0}^6 r^{8-6 \beta }}{1-6 \beta
   }\nonumber\\&&-\frac{256 \alpha  \beta ^5 (2 \beta -3) e^{6 {u_0}} \left(102 \beta ^2-3 \beta  (8 r+89)+28 (r+6)\right) r^{6 \beta }}{(6 \beta -7) {w_0}^5}+\frac{9 \beta ^3 e^{-2
   {u_0}} {w_0}^3 r^{5-2 \beta }}{\beta +1}+12 \beta  r^6 {w_0}\Big)
\end{eqnarray}
\par
$\bullet$ Fourth case\par

In this case, we assume the parameters $w(r)$ and $k(r)$ of the line element as 
\begin{eqnarray}
w(r)=\sqrt{2}\,,\quad k(r)=\ln{\left[w(r)\right]}\,.
\end{eqnarray}
As we performed in the previous case, we need to find a suitable expression for the line element parameter $u(r)$. To do so, we assume the following relation between $G(r)$ and $R(r)$ as
\begin{eqnarray}
G(r)=b_1(r)R(r)+b_2(r)\,\,,
\end{eqnarray}
such that action algebraic function becomes 
\begin{eqnarray}
R+f(G)=\left[1+2\alpha b_1b_2\right]R+\alpha \left(b_1^2R^2+b_2^2\right)\,\,.
\end{eqnarray}
Now, by introducing an auxiliary function $\psi$ to be identified to the radial coordinate $r$ ($\psi=r$), one can rewrite the action algebraic function as
\begin{eqnarray}
R+f(G)=B_1(\psi)R+B_2(\psi)R^2+V(\psi)\,\,,
\end{eqnarray}
such that $G(r)$ and $R(r)$ become
\begin{eqnarray}
G(r)=2e^{4u(r)}u'^2(r)\left[2u'^2+3u''(r)\right]\,\,,\label{gau}\\
R(r)=e^{2u(r)}\left[u'^2(r)-u''(r)\right]\,\,.\label{curv}
\end{eqnarray}
Once again, relation is required between $G$ and $R$ in order to find a suitable expression for the line element parameter $u(r)$. To this end we assume the following relation
\begin{eqnarray}
G(r)=4e^{2u(r)}u'^2R(r)\,\,\,,
\end{eqnarray}
from which, using (\ref{gau}) and (\ref{curv}), one gets 
\begin{eqnarray}
u(r)=\frac{1}{2}\ln{\left[\frac{2}{3}\sqrt{\frac{2}{5}}r^{3/2}\right]}
\end{eqnarray}

By making use of these solutions, one gets the energy density and pressures as
\begin{eqnarray}
\rho(r)&=&\frac{549\alpha-220\sqrt{10}r^{3/2}}{1600\kappa^2r^2}\nonumber\\
p_r(r)&=&\frac{3\left(81\alpha-20\sqrt{10}r^{3/2}\right)}{1600\kappa^2r^2}\nonumber\\
p_{\varphi}(r)&=&\frac{3\left(57\alpha+20\sqrt{10}r^{3/2}\right)}{1600\kappa^2r^2}\nonumber\\
p_z(r)&=&\frac{3\left[20\sqrt{10}r^{3/2}+\alpha\left(57+96r\right)\right]}{1600\kappa^2 r^2}
\end{eqnarray}
Now the metric reads
\begin{eqnarray}
ds^2=x^{3/2}d\eta^2-x^{-3/2}\Big(dx^2+d\xi^2+d\psi^2\Big),\\\nonumber x=\sqrt{3\sqrt{5/2}}r,\ \ \eta=4\sqrt[11/4]{\frac{1}{3}\sqrt{2/5}},\ \ \xi=\sqrt[19/8]{3\sqrt{5/2}}z,\ \ \psi=\sqrt[19/8]{3\sqrt{5/2}}.
\end{eqnarray}
This is a very important result and it shows a scale independent solution with hypers-calling parameter $\alpha=0$. The observers in all points of this space time measures the physical lengths without any need to including the redshift factors, because this solution is scale independent.
For this case the mass reads
\begin{eqnarray}
m=\frac{27 \sqrt{5} \left(9-160 \sqrt{2} {w_0}^4\right) {w_0}^2}{448 r^{7/2}}+\frac{3}{\sqrt{2} r}.
\end{eqnarray}
Here for $r\rightarrow 0$, $m\rightarrow\infty$.

\par
$\bullet$ The fifth case \par
In this case the parameters of the line element are 
\begin{eqnarray}
w(r)=w_0+w_1r\,,\quad k(r)=\ln{\left[w(r)\right]}\,,\quad u(r)=u_0\ln{\left[w(r)\right]}.
\end{eqnarray}
The Gauss-Bonnet invariant term and the curvature scalar read
\begin{eqnarray}
G(r)&=&16u_0w_1^4\left(u_0-2\right)\left(u_0-1\right)^2\left(w_0+w_1r\right)^{4(u_0-2)}\,\,,\\
R(r)&=&2w_1^2\left(u_0^2-1\right)\left(w_0+w_1r\right)^{2(u_0-2)}\,\,.
\end{eqnarray}
Therefore, the corresponding energy density and pressures read
\begin{eqnarray}
\rho(r)=-\frac{w_1^2}{\kappa^2}\left(u_0-1\right)\left(w_0+w_1r\right)^{2(u_0-8)}\Big\{
\left(u_0+1\right)\Big(w_0^{12}+12w_1w_0^{11}r+66w_1^2w_0^{10}r^2+220w_0^9w_1^3r^3
\nonumber\\
495w_0^8w_1^4r^4+792w_0^7w_1^5r^5+924w_0^6w_1^6r^6+792w_0^5w_1^7r^7+495w_0^4w_1^8r^8+
220w_0^3w_1^9r^9\nonumber\\
66w_0^2w_1^{10}r^{10}+12w_0w_1^{11}r^{11}\Big)+w_1^6\Big[w_1^6(1+u_0)r^{12}\nonumber\\
+256\alpha u_0(u_0-2)^2(u_0-1)^3(19u_0-48)(w_0+w_1r)^{6u_0}\Big]\Big\}
\end{eqnarray}
\begin{eqnarray}
p_r(r)=-\frac{w_1^2}{\kappa^2}\left(u_0-1\right)\left(w_0+w_1r\right)^{2(u_0-8)}\Big\{
\left(u_0+1\right)\Big(w_0^{12}+12w_1w_0^{11}r+66w_1^2w_0^{10}r^2+220w_0^9w_1^3r^3
\nonumber\\
495w_0^8w_1^4r^4+792w_0^7w_1^5r^5+924w_0^6w_1^6r^6+792w_0^5w_1^7r^7+495w_0^4w_1^8r^8+
220w_0^3w_1^9r^9\nonumber\\
66w_0^2w_1^{10}r^{10}+12w_0w_1^{11}r^{11}\Big)+w_1^6\Big[w_1^6(1+u_0)r^{12}\nonumber\\
-256\alpha u_0^2(u_0-2)^2(11u_0^3-9u_0^2+9u_0-11)(w_0+w_1r)^{6u_0}\Big]\Big\}
\end{eqnarray}
\begin{eqnarray}
p_{\varphi}(r)=-\frac{w_1^2}{\kappa^2}\left(u_0-1\right)\left(w_0+w_1r\right)^{2(u_0-8)}\Big\{
\left(u_0+1\right)\Big(w_0^{12}+12w_1w_0^{11}r+66w_1^2w_0^{10}r^2+220w_0^9w_1^3r^3
\nonumber\\
495w_0^8w_1^4r^4+792w_0^7w_1^5r^5+924w_0^6w_1^6r^6+792w_0^5w_1^7r^7+495w_0^4w_1^8r^8+
220w_0^3w_1^9r^9\nonumber\\
66w_0^2w_1^{10}r^{10}+12w_0w_1^{11}r^{11}\Big)+w_1^6\Big[w_1^6(1+u_0)r^{12}\nonumber\\
-256\alpha u_0^2(2-3u_0+u_0^2)^2(29u_0-53)(w_0+w_1r)^{6u_0}\Big]\Big\}
\end{eqnarray}
\begin{eqnarray}
p_z(r)=-\frac{w_1^2}{\kappa^2}\left(u_0-1\right)\left(w_0+w_1r\right)^{2(u_0-8)}\Big\{
\left(u_0+1\right)\Big(w_0^{12}+12w_1w_0^{11}r+66w_1^2w_0^{10}r^2+220w_0^9w_1^3r^3
\nonumber\\
495w_0^8w_1^4r^4+792w_0^7w_1^5r^5+924w_0^6w_1^6r^6+792w_0^5w_1^7r^7+495w_0^4w_1^8r^8+
220w_0^3w_1^9r^9\nonumber\\
66w_0^2w_1^{10}r^{10}+12w_0w_1^{11}r^{11}\Big)+2048\alpha w_0w_1^5u_0^3(u_0-1)(u_0-1)(w_0+w_1r)^{6u_0}\nonumber\\
+w_1^6\Big[w_1^6(1+u_0)r^{12}+2048\alpha u_0^3r(u_0-1)(u_0-2)^2(w_0+w_1r)^{6u_0}\nonumber\\-256\alpha u_0^2(u_0-2)^2(29u_0^3-103u_0^3+127u_0-53)(w_0+w_1r)^{6u_0}\Big]\Big\}
\end{eqnarray}
This case is analogue to the LC family in GB gravity, which has been discussed recently \cite{cylinder-gb1}. The mass reads
\begin{eqnarray}
m&=&-\frac{128 \alpha  ({u_0}-2)^2 ({u_0}-1)^2 {u_0}^2 ({u_0} (5 {u_0}-17)+11) {w_1}^7 (r {w_1}+{w_0})^{2 {u_0}-7}}{2
   {u_0}-7}\nonumber\\&&+\frac{{u_0} ({u_0} (3 {u_0}-5)+3) {w_1}^2 (r {w_1}+{w_0})^{1-2 {u_0}}}{2 {u_0}-1}+\frac{8 (r {w_1}+{w_0})^{7-6
   {u_0}}}{7 {w_1}-6 {u_0} {w_1}}+\frac{\alpha\Delta_2}{3 (-11 + 6 u_0)}\\
   \Delta_2&=&256 ({u_0}-2) ({u_0}-1)^2 {u_0} {w_1}^6 (r {w_1}+{w_0})^{6 ({u_0}-2)} ({w_1} ({u_0} (12 r ((3-2 {u_0}) {u_0}+2)\nonumber\\&&+{u_0} ({u_0}
   (102 {u_0}-655)+1308)-969)+264)+12 {u_0} ((3-2 {u_0}) {u_0}+2) {w_0}).
\end{eqnarray}

\par
\section{Solutions for the model $f(G)=\alpha G^3$}
In this case, like the previous solutions, the model is viable. So as a particular solution for the model, the expressions for the line element parameters are
\begin{eqnarray}
w(r)=w_0\,,\quad k(r)=\ln{\left(w(r)\right)}\,,\quad u(r)=\ln{\left[w(r)+u_0r^2\right]}\,\,.
\end{eqnarray}
Therefore, the Gauss-Bonnet invariant remains the same in (\ref{GB26}). But the energy density and pressure change and are written as
\begin{eqnarray}
\rho(r)=-\frac{4u_0}{\kappa^2w_0^{12}}\Big\{5373952\alpha u_0^{11}r^{12}+29097984\alpha u_0^{10}w_0r^{10}+56623104\alpha u_0^9w_0^2r^8\nonumber\\
+49545216\alpha u_0^8w_0^3r^6+17694720\alpha u_0^7w_0^4r^4+2u_0w_0^{10}r^2-w_0^{11}\Big\}
\end{eqnarray}
\begin{eqnarray}
p_r(r)=\frac{4u_0^2r^2}{\kappa^2w_0^{12}}\Big\{4587520\alpha u_0^{10}r^{10}
+24772608\alpha u_0^9w_0\nonumber\\
38928284\alpha u_0^8w_0^2r^6+17694720\alpha u_0^7w_0^3r^4-w_0^{10}\Big\}
\end{eqnarray}
\begin{eqnarray}
p_{\varphi}(r)=-\frac{4u_0^2r^2}{\kappa^2w_0^{12}}\Big\{8781824\alpha u_0^{10}r^{10}+48758784\alpha u_0^{9}w_0r^{8}+92012544\alpha u_0^{8}w_0^2r^6\nonumber\\
+70778880\alpha u_0^7w_0^3r^4+17694720\alpha u_0^6w_0^4r^2-w_0^{10}\Big\}
\end{eqnarray}
\begin{eqnarray}
p_z(r)=\frac{4u_0^2r^2}{\kappa^2w_0^{12}}\Big\{-8781824\alpha u_0^{10}r^{10}+1572864 \alpha u_0^{10}r^{11}-48758784\alpha u_0^9w_0r^8\nonumber\\
-48758784\alpha u_0^9w_0r^8+10223616\alpha u_0^9w_0r^9-92012544\alpha u_0^8w_0^2r^6\nonumber\\
+22806528\alpha u_0^8w_0^2r^7-70778880\alpha u_0^7w_0^3r^4+21233664\alpha u_0^7w_0^3r^5\nonumber\\
-17694720\alpha u_0^6w_0^4+7077888\alpha u_0^6w_0^4r^3+w_0^{10}\Big\}
\end{eqnarray}
So, the line element reads
\begin{eqnarray}
ds^2=(1+(\frac{r}{r_0}))^2d\eta^2-\frac{dr^2+dz^2+d\varphi^2}{(1+(\frac{r}{r_0}))^2},\ \ r_0=\sqrt{\frac{w_0}{u_0}},\eta=w_0t.
\end{eqnarray}
This metric is conform-flat-stationary metric of one parameter family of the exact solutions. The mass reads
\begin{eqnarray}
m=\frac{4\mathcal{C}}{w_0^9},
\end{eqnarray}
where
\begin{eqnarray}
\mathcal{C}&=&\frac{786432}{5} \alpha  r^{10} {u_0}^{10}-\frac{9306112}{9} \alpha  r^9 {u_0}^{10}+884736 \alpha  r^8 {u_0}^9 {w_0}-4587520 \alpha  r^7 {u_0}^9
   {w_0}+1179648 \alpha  r^6 {u_0}^8 {w_0}^2\nonumber\\&&-\frac{29097984}{5} \alpha  r^5 {u_0}^8 {w_0}^2-\frac{5242880}{3} \alpha  r^3 {u_0}^7 {w_0}^3+\frac{r
   {w_0}^5 (262144 \alpha  {u_0}^6 (141 {w_0}-256)+(128 {u_0}+63) {w_0}^5)}{128 (r^2 {u_0}+{w_0})}\nonumber\\&&+\frac{r {w_0}^{10}
   (196608 \alpha  {u_0}^6+{w_0}^4)}{5 (r^2 {u_0}+{w_0})^5}+\frac{3 {w_0}^9 ({w_0}^4 (3 r-10 {u_0})-1048576 \alpha  r
   {u_0}^6)}{40 (r^2 {u_0}+{w_0})^4}\nonumber\\&&+\frac{{w_0}^8 ({w_0}^4 (21 r+80 {u_0})-8650752 \alpha  r {u_0}^6)}{80 (r^2
   {u_0}+{w_0})^3}+\frac{3 r {w_0}^7 (262144 \alpha  {u_0}^6+7 {w_0}^4)}{64 (r^2 {u_0}+{w_0})^2}\nonumber\\&&-16384 \alpha  r
   {u_0}^6 {w_0}^4 (33 {w_0}-256)+\frac{7 {w_0}^{9/2} (262144 \alpha  {u_0}^6 (27 {w_0}-256)+9 {w_0}^5) \tan ^{-1}(\frac{r
   \sqrt{{u_0}}}{\sqrt{{w_0}}})}{128 \sqrt{{u_0}}}
\end{eqnarray}

\par
\section{Model $f(G)=\alpha_3 \exp{(G)}$}
This later case is analogue of the exponential  $f(R)$-gravity model which reproduces the current epoch of the accelerated expansion of the universe and also the de-Sitter final stage as well \cite{sergei4}. By replacing the curvature role by $G$, it can be used as an alternative viable model with stable attractors in $f(G)$ gravity. So, we apply it to construct the exact solutions.
For such more complicated model, we find solution for the line element, i.e, obtaining the expressions of $w(r)$, $u(r)$ and $k(r)$. For this we choose the parameters as 
\begin{eqnarray}
u(r)=u_0\ln{\left(w(r)\right)}\,\,,\quad k(r)=k_0\ln{\left(w(r)\right)}\,,\quad w(r)=r\,\,.\label{lineelements}
\end{eqnarray}
Therefore, the Gauss-Bonnet invariant and the curvature scalar read
\begin{eqnarray}
G(r)&=&16u_0(u_0-1)(k_0-u_0)(k_0-u_0+1)r^{4(u_0-k_0-1)}\,\,,\\
R(r)&=&2(u_0^2-k_0)r^{2(u_0-k_0-1)}\,\,.
\end{eqnarray}
Still making use of (\ref{lineelements}), one gets the energy density and pressures as
\begin{eqnarray}
\rho(r)=\frac{1}{\kappa^2}\Big\{(k_0-u_0^2)r^{2(5+5k_0+u_0)}-\alpha_3u_0(k_0-u_0)(u_0-1)(1+k_0-u_0)\exp{\left(16r^{4(u_0-k_0-1)}\right)}\times\nonumber\\
\Big[r^{12(1+k_0)}-16u_0(u_0-1)(k_0-u_0)(1+k_0-u_0)r^{4(u_0+2k_0+2)}\nonumber\\
-512u_0(u_0-1)^2(k_0-u_0)^2(1+k_0-u_0)^2(6+6k_0-5u_0)r^{4(1+k_0+2u_0)}\nonumber\\
+32768u_0^2(u_0-1)^3(u_0-k_0)^3(1+k_0-u_0)^4r^{12u_0}\Big]\Big\}r^{-12(1+k_0)}\,\,,
\end{eqnarray}
\begin{eqnarray}
p_r(r)=\frac{1}{\kappa^2}\Big\{\alpha_3u_0(u_0-1)(k_0-u_0)(1+k_0-u_0)\exp{\left(16r^{4(u_0-k_0-1)}\right)}+(k_0-u_0^2)r^{2(u_0-k_0-1)}\nonumber\\
-16\alpha_3u_0^2(u_0-1)^2(k_0-u_0)^2(1+k_0-u_0)^2\exp{\left(16r^{4(u_0-k_0-1)}\right)}r^{4(u_0-2k_0-2)}\Big[r^{4(1+k_0)}-96u_0r^{4u_0}\Big(k_0^2(u_0-1)\nonumber\\
-k_0(2u_0^2-u_0+1)+u_0(u_0^2-1)\Big)\Big]\Big\}\,\,,
\end{eqnarray}

\begin{eqnarray}
p_{\varphi}(r)&=&\frac{1}{\kappa^2}\Big\{\alpha_3u_0(u_0-1)(k_0-u_0)(1+k_0-u_0)\exp{\left(16r^{4(u_0-k_0-1)}\right)}+(u_0^2-k_0)r^{2(u_0-k_0-1)}\nonumber\\
&-&16\alpha_3u_0^2(u_0-1)^2(k_0-u_0)^2(1+k_0-u_0)^2\exp{\left(16r^{4(u_0-k_0-1)}\right)}r^{4(u_0-3-3k_0)}\Big[r^{8(1+k_0)}\nonumber\\
&+&2048r^{8u_0}u_0^2(k_0-u_0)^2(u_0-1)(u_0-k_0-1)^3\nonumber\\
&+&32r^{4(1+k_0+u_0)}u_0\Big(-6k_0^3+7u_0(u_0-1)^2+k_0^2(19u_0-13)
-k_0(20u_0^2-27u_0+7)\Big)\Big]\Big\}\,\,,
\end{eqnarray}
\begin{eqnarray}
p_z&=&\frac{1}{\kappa^2}\Big\{\alpha_3u_0(u_0-1)(k_0-u_0)(1+k_0-u_0)\exp{\left(16r^{4(u_0-k_0-1)}\right)}+(u_0^2-k_0)r^{2(u_0-k_0-1)}\nonumber\\
&-&16\alpha_3u_0^2(u_0-1)^2(k_0-u_0)^2(1+k_0-u_0)^2r^{4(u_0-3k_0-3)}\exp{\left(16r^{4(u_0-k_0-1)}\right)}\Big[r^{8(1+k_0)}\nonumber\\
&-&64u_0^2(u_0-k_0-1)r^{5+4(k_0+u_0)}+2048r^{8u_0}u_0^2(u_0-k_0)(u_0-1)^2(u_0-k_0-1)^3\nonumber\\
&+&32r^{4(1+k_0+u_0)}u_0\Big( -6+17u_0-18u_0^2+7u_0^3+7k_0^2(u_0-1)-k_0(14u_0^2-25u_0+13)               \Big)\Big]\Big\}\,.
\end{eqnarray}
On the other hand, by setting $u_0=M$ and $k_0=M^2$, the parameters are written 
\begin{eqnarray}
k(r)=M^2\ln{(r)}\,,\quad u(r)=M\ln{(r)}\,\,,\quad w(r)=r\,\,,
\end{eqnarray} 
such that the curvature scalar vanishes. In this case, the Gauss-Bonnet invariant reads
\begin{eqnarray}
G=16M^2(M^2-1)(1-M+M^2)r^{-4(M^2-M+1)}\,\,\,.
\end{eqnarray}
Thereby, the energy density and pressures become
\begin{eqnarray}
\rho(r)=\frac{r^{-12(1+M^2)}}{\kappa^2}\Big\{32768M^5(M-1)^6\left[M(M-1)+1\right]r^{12M}\nonumber\\
+512M^3(M-1)^4\left[M(M-1)+1\right]^2\left[6+M(6M-5)\right]r^{4(1+M)^2}-r^{12(1+M^2)}\nonumber\\
-16M^2(M-1)^2\left[1+M(M-1)\right]r^{4(2M^2+M+2)}\Big\}f(G)
\end{eqnarray}
\begin{eqnarray}
p_r(r)=\frac{1}{\kappa^2}\Big\{1-16M^2(M-1)^2\left[M(M-1)+1\right]\Big[-96M^2\left[M(M-2)-1\right]
\times\nonumber\\
\left[1+M(M-1)\right]r^{4M}+r^{4(M^2+1)}\Big]r^{-4(2M^2-M+2)}\Big\}f(G)
\end{eqnarray}
\begin{eqnarray}
p_{\varphi}(r)=\frac{1}{\kappa^2}\Big\{1-16M^2(M-1)^2\left[M(M-1)+1\right]\times\nonumber\\
\Big[-2048(M-1)^3M^4\left[M(M-1)+1\right]^3r^{8M}-32M^2(M-1)\left[M(M-1)+1\right]
\times\nonumber\\
\left[M(6M-7)+7\right]\left[M(6M-7)+7\right]r^{4(1+M+M^2)}+r^{8(1+M^2)}\Big]r^{-4(3M^2-M+3)}\Big\}f(G)
\end{eqnarray}
\begin{eqnarray}
p_z(r)=\frac{1}{\kappa^2}\Big\{1-16M^2(M-1)^2\left[M(M-1)+1\right]\times\nonumber\\
\Big[2048M^3(M-1)^3\left[M(M-1)+1\right]^3+r^{8+8M(M-1)}\nonumber\\
+32Mr^{4+4M(M-1)}\left[M(M-1)+1\right]\Big(-6+M\left[11+7M(M-2)+2r\right]\Big)\Big]
r^{-12[1+M(M-1)]}\Big\}f(G)
\end{eqnarray}
This is exactly the LC family but with $G\neq0$ in $f(G)$ \cite{cylinder-gb1}.

\section{Analysis of the energy conditions}
In this section we will do the full analysis on the NEC for all families of the solutions which have been presented in the previous sections. We list the cases below:
\subsection{Model $f(G)=\alpha G^2$}
\par
$\bullet$ First case\par
For the first NEC for radial pressure $p_r$ we will have
\begin{eqnarray}
 &&\frac{2 u_1 w_1^2 \mathcal{P}}{(w_0 + r w_1)^15 (u_1 + w_0 + r w_1)^4}\geq0
 \end{eqnarray}
 Where in it
 \begin{eqnarray}
\mathcal{P}&=&3 (w_0 + r w_1)^16 + 14784 u1^10 w_1^6 \alpha + 
 192 u1^9 w_1^5 (w_0 + r w_1) (7 w_0 + (580 + 7 r) w_1) \alpha \nonumber\\&&+ 
 576 u1^8 w_1^5 (w_0 + r w_1)^2 (19 w_0 + (642 + 19 r) w_1) \alpha + 
 64 u1^7 w_1^5 (w_0 + r w_1)^3 (609 w_0 + (11215 + 609 r) w_1) \alpha\nonumber\\&& + 
 64 u1^6 w_1^5 (w_0 + r w_1)^4 (1239 w_0 + (14302 + 1239 r) w_1) \alpha +
  u1 (w_0 + r w_1)^9 (13 (w_0 + r w_1)^6 \nonumber\\&&+ 
    1536 w_1^5 (w_0 + (7 + r) w_1) \alpha) + 
 2 u1^2 (w_0 + r w_1)^8 (11 (w_0 + r w_1)^6 + 
    288 w_1^5 (21 w_0 + (131 + 21 r) w_1) \alpha)\nonumber\\&& + 
 u1^5 (w_0 + r w_1)^5 ((w_0 + r w_1)^6 - 
    24 w_1^5 (3 w_0 (-1400 + w_1) + (-34192 + 
          3 r (-1400 + w_1)) w_1) \alpha)\nonumber\\&& + 
 u1^4 (w_0 + r w_1)^6 (7 (w_0 + r w_1)^6 - 
    64 w_1^5 (3 w_0 (-427 + w_1) + (-8374 + 
          3 r (-427 + w_1)) w_1) \alpha) \nonumber\\&&+ 
 2 u_1^3 (w_0 + r w_1)^7 (9 (w_0 + r w_1)^6 - 
    32 w_1^5 (w_0 (-651 + 2 w_1) + w_1 (-3919 - 651 r + 2 r w_1)) \alpha)
\end{eqnarray}
Further for the azimuthal pressure we have
\begin{eqnarray}
&&\frac{\mathcal{Q}}{(w_0 + r w_1)^15 (u_1 + w_0 + r w_1)^4}\geq0\,\,.
\end{eqnarray}
Here
\begin{eqnarray}
\mathcal{Q}&=&-r^{16} {w_1}^{19}+8 r^{18} {w_1}^{18}+2 r^{16} {u_1} {w_1}^{18}-2
   r^{15} {u_1} {w_1}^{18}-16 r^{15} {w_0} {w_1}^{18}+8 r^{15}
   {u_1}^2 {w_1}^{17}\nonumber\\&&-4 r^{14} {u_1}^2 {w_1}^{17}-120 r^{14}
   {w_0}^2 {w_1}^{17}+144 r^{17} {w_0} {w_1}^{17}+32 r^{15} {u_1}
   {w_0} {w_1}^{17}-30 r^{14} {u_1} {w_0} {w_1}^{17}+12 r^{14}
   {u_1}^3 {w_1}^{16}\nonumber\\&&-3 r^{13} {u_1}^3 {w_1}^{16}-560 r^{13}
   {w_0}^3 {w_1}^{16}+1224 r^{16} {w_0}^2 {w_1}^{16}+240 r^{14}
   {u_1} {w_0}^2 {w_1}^{16}\nonumber\\&&-210 r^{13} {u_1} {w_0}^2
   {w_1}^{16}+120 r^{14} {u_1}^2 {w_0} {w_1}^{16}\nonumber\\&&-56 r^{13}
   {u_1}^2 {w_0} {w_1}^{16}+8 r^{13} {u_1}^4 {w_1}^{15}-1820
   r^{12} {w_0}^4 {w_1}^{15}+6528 r^{15} {w_0}^3 {w_1}^{15}+1120
   r^{13} {u_1} {w_0}^3 {w_1}^{15}\nonumber\\&&-910 r^{12} {u_1} {w_0}^3
   {w_1}^{15}+840 r^{13} {u_1}^2 {w_0}^2 {w_1}^{15}\nonumber\\&&-364 r^{12}
   {u_1}^2 {w_0}^2 {w_1}^{15}+168 r^{13} {u_1}^3 {w_0}
   {w_1}^{15}-39 r^{12} {u_1}^3 {w_0} {w_1}^{15}\nonumber\\&&+2 r^{12} {u_1}^5
   {w_1}^{14}-4368 r^{11} {w_0}^5 {w_1}^{14}+24480 r^{14} {w_0}^4
   {w_1}^{14}+3640 r^{12} {u_1} {w_0}^4 {w_1}^{14}-2730 r^{11}
   {u_1} {w_0}^4 {w_1}^{14}\nonumber\\&&+3640 r^{12} {u_1}^2 {w_0}^3
   {w_1}^{14}-1456 r^{11} {u_1}^2 {w_0}^3 {w_1}^{14}+1092 r^{12}
   {u_1}^3 {w_0}^2 {w_1}^{14}\nonumber\\&&-234 r^{11} {u_1}^3 {w_0}^2
   {w_1}^{14}+104 r^{12} {u_1}^4 {w_0} {w_1}^{14}\nonumber\\&&-8008 r^{10}
   {w_0}^6 {w_1}^{13}+68544 r^{13} {w_0}^5 {w_1}^{13}+8736 r^{11}
   {u_1} {w_0}^5 {w_1}^{13}\nonumber\\&&-6006 r^{10} {u_1} {w_0}^5
   {w_1}^{13}+10920 r^{11} {u_1}^2 {w_0}^4 {w_1}^{13}\nonumber\\&&-4004 r^{10}
   {u_1}^2 {w_0}^4 {w_1}^{13}+4368 r^{11} {u_1}^3 {w_0}^3
   {w_1}^{13}-858 r^{10} {u_1}^3 {w_0}^3 {w_1}^{13}\nonumber\\&&+624 r^{11}
   {u_1}^4 {w_0}^2 {w_1}^{13}+24 r^{11} {u_1}^5 {w_0}
   {w_1}^{13}\nonumber\\&&-11440 r^9 {w_0}^7 {w_1}^{12}+148512 r^{12} {w_0}^6
   {w_1}^{12}+16016 r^{10} {u_1} {w_0}^6 {w_1}^{12}-10010 r^9
   {u_1} {w_0}^6 {w_1}^{12}+24024 r^{10} {u_1}^2 {w_0}^5
   {w_1}^{12}\nonumber\\&&-8008 r^9 {u_1}^2 {w_0}^5 {w_1}^{12}+12012 r^{10}
   {u_1}^3 {w_0}^4 {w_1}^{12}-2145 r^9 {u_1}^3 {w_0}^4
   {w_1}^{12}+2288 r^{10} {u_1}^4 {w_0}^3 {w_1}^{12}\nonumber\\&&+132 r^{10}
   {u_1}^5 {w_0}^2 {w_1}^{12}-12870 r^8 {w_0}^8 {w_1}^{11}+254592
   r^{11} {w_0}^7 {w_1}^{11}+22880 r^9 {u_1} {w_0}^7
   {w_1}^{11}-12870 r^8 {u_1} {w_0}^7 {w_1}^{11}\nonumber\\&&+40040 r^9
   {u_1}^2 {w_0}^6 {w_1}^{11}\nonumber\\&&-12012 r^8 {u_1}^2 {w_0}^6
   {w_1}^{11}+24024 r^9 {u_1}^3 {w_0}^5 {w_1}^{11}-3861 r^8
   {u_1}^3 {w_0}^5 {w_1}^{11}\nonumber\\&&+5720 r^9 {u_1}^4 {w_0}^4
   {w_1}^{11}+440 r^9 {u_1}^5 {w_0}^3 {w_1}^{11}-11440 r^7
   {w_0}^9 {w_1}^{10}+350064 r^{10} {w_0}^8 {w_1}^{10}\nonumber\\&&+25740 r^8
   {u_1} {w_0}^8 {w_1}^{10}-12870 r^7 {u_1} {w_0}^8
   {w_1}^{10}+51480 r^8 {u_1}^2 {w_0}^7 {w_1}^{10}\nonumber\\&&-13728 r^7
   {u_1}^2 {w_0}^7 {w_1}^{10}+36036 r^8 {u_1}^3 {w_0}^6
   {w_1}^{10}\nonumber\\&&-5148 r^7 {u_1}^3 {w_0}^6 {w_1}^{10}+10296 r^8
   {u_1}^4 {w_0}^5 {w_1}^{10}+990 r^8 {u_1}^5 {w_0}^4
   {w_1}^{10}\nonumber\\&&-8008 r^6 {w_0}^{10} {w_1}^9+388960 r^9 {w_0}^9
   {w_1}^9\nonumber\\&&+22880 r^7 {u_1} {w_0}^9 {w_1}^9-10010 r^6 {u_1}
   {w_0}^9 {w_1}^9+51480 r^7 {u_1}^2 {w_0}^8 {w_1}^9-12012 r^6
   {u_1}^2 {w_0}^8 {w_1}^9\nonumber\\&&+41184 r^7 {u_1}^3 {w_0}^7
   {w_1}^9-5148 r^6 {u_1}^3 {w_0}^7 {w_1}^9+13728 r^7 {u_1}^4
   {w_0}^6 {w_1}^9+1584 r^7 {u_1}^5 {w_0}^5 {w_1}^9\nonumber\\&&-4368 r^5
   {w_0}^{11} {w_1}^8+350064 r^8 {w_0}^{10} {w_1}^8\nonumber\\&&+16016 r^6
   {u_1} {w_0}^{10} {w_1}^8-6006 r^5 {u_1} {w_0}^{10}
   {w_1}^8+40040 r^6 {u_1}^2 {w_0}^9 {w_1}^8-8008 r^5 {u_1}^2
   {w_0}^9 {w_1}^8+36036 r^6 {u_1}^3 {w_0}^8 {w_1}^8\nonumber\\&&-3861 r^5
   {u_1}^3 {w_0}^8 {w_1}^8+13728 r^6 {u_1}^4 {w_0}^7
 {w_1}^8+1848 r^6 {u_1}^5 {w_0}^6 {w_1}^8+\mathcal{X}
 \end{eqnarray}
 \begin{eqnarray}
 \mathcal{X}&=&-1820 r^4 {w_0}^{12}
   {w_1}^7+254592 r^7 {w_0}^{11} {w_1}^7+8736 r^5 {u_1}
   {w_0}^{11} {w_1}^7-2730 r^4 {u_1} {w_0}^{11} {w_1}^7\nonumber\\&&+24024 r^5
   {u_1}^2 {w_0}^{10} {w_1}^7-4004 r^4 {u_1}^2 {w_0}^{10}
   {w_1}^7+24024 r^5 {u_1}^3 {w_0}^9 {w_1}^7\nonumber\\&&-2145 r^4 {u_1}^3
   {w_0}^9 {w_1}^7+10296 r^5 {u_1}^4 {w_0}^8 {w_1}^7+1584 r^5
   {u_1}^5 {w_0}^7 {w_1}^7+16 {u_1}^2 (1848 {w_1}
   {u_1}^9\nonumber\\&&+24 ({w_0}+r {w_1}) (7 {w_0}+(7 r+559) {w_1})
   {u_1}^8+72 ({w_0}+r {w_1})^2 (19 {w_0}+(19 r+585) {w_1})
   {u_1}^7\nonumber\\&&+8 ({w_0}+r {w_1})^3 (609 {w_0}+(609 r+9325) {w_1})
   {u_1}^6\nonumber\\&&+8 ({w_0}+r {w_1})^4 (1239 {w_0}+(1239 r+10198) {w_1})
   {u_1}^5-3 ({w_0}+r {w_1})^5 ({w_0} (59 {w_1}-4200)+{w_1}
   (59 {w_1} r\nonumber\\&&-4200 r-18952)) {u_1}^4-({w_0}+r {w_1})^6 (183
   {w_0} (3 {w_1}-56)+{w_1} (183 r (3 {w_1}-56)-25448)) {u_1}^3\nonumber\\&&-8
   ({w_0}+r {w_1})^7 ({w_0} (68 {w_1}-651)+{w_1}\nonumber\\&& (68 {w_1}
   r-651 r-931)) {u_1}^2-21 ({w_0}+r {w_1})^8 ({w_0} (7
   {w_1}-72)+{w_1} (7 r {w_1}\nonumber\\&&-72 (r+1))) {u_1}+24 ({w_0}+r
   {w_1})^9 ({w_0} ({w_1}+8)+{w_1} (r ({w_1}+8)+8))) \alpha
    {w_1}^7-560 r^3 {w_0}^{13} {w_1}^6\nonumber\\&&+148512 r^6 {w_0}^{12}
   {w_1}^6+3640 r^4 {u_1} {w_0}^{12} {w_1}^6\-910 r^3 {u_1}
   {w_0}^{12} {w_1}^6\nonumber\\&&+10920 r^4 {u_1}^2 {w_0}^{11} {w_1}^6-1456
   r^3 {u_1}^2 {w_0}^{11} {w_1}^6+12012 r^4 {u_1}^3 {w_0}^{10}
   {w_1}^6-858 r^3 {u_1}^3 {w_0}^{10} {w_1}^6\nonumber\\&&+5720 r^4 {u_1}^4
   {w_0}^9 {w_1}^6+990 r^4 {u_1}^5 {w_0}^8 {w_1}^6-120 r^2
   {w_0}^{14} {w_1}^5+68544 r^5 {w_0}^{13} {w_1}^5\nonumber\\&&+1120 r^3 {u_1}
   {w_0}^{13} {w_1}^5-210 r^2 {u_1} {w_0}^{13} {w_1}^5\nonumber\\&&+3640 r^3
   {u_1}^2 {w_0}^{12} {w_1}^5-364 r^2 {u_1}^2 {w_0}^{12}
   {w_1}^5+4368 r^3 {u_1}^3 {w_0}^{11} \nonumber\\&&{w_1}^5-234 r^2 {u_1}^3
   {w_0}^{11} {w_1}^5+2288 r^3 {u_1}^4 {w_0}^{10} {w_1}^5\nonumber\\&&+440 r^3
   {u_1}^5 {w_0}^9 {w_1}^5\nonumber\\&&-16 r {w_0}^{15} {w_1}^4+24480 r^4
   {w_0}^{14} {w_1}^4+240 r^2 {u_1} {w_0}^{14} {w_1}^4-30 r
   {u_1} {w_0}^{14} {w_1}^4+840 r^2 {u_1}^2 {w_0}^{13}
   {w_1}^4\nonumber\\&&-56 r {u_1}^2 {w_0}^{13} {w_1}^4+1092 r^2 {u_1}^3
   {w_0}^{12} {w_1}^4\nonumber\\&&-39 r {u_1}^3 {w_0}^{12} {w_1}^4+624 r^2
   {u_1}^4 {w_0}^{11} {w_1}^4\nonumber\\&&+132 r^2 {u_1}^5 {w_0}^{10}
  {w_1}^4-{w_0}^{16}{w_1}^3+6528 r^3{w_0}^{15}{w_1}^3+32 r
  {u_1}{w_0}^{15}{w_1}^3\nonumber\\&&-2{u_1}{w_0}^{15}{w_1}^3+120 r
  {u_1}^2{w_0}^{14}{w_1}^3-4{u_1}^2{w_0}^{14}
  {w_1}^3+168 r{u_1}^3{w_0}^{13}{w_1}^3-3{u_1}^3
  {w_0}^{13}{w_1}^3\nonumber\\&&+104 r{u_1}^4{w_0}^{12}{w_1}^3+24 r
  {u_1}^5{w_0}^{11}{w_1}^3\nonumber\\&&+1224 r^2{w_0}^{16}{w_1}^2+2
  {u_1}{w_0}^{16}{w_1}^2+8{u_1}^2{w_0}^{15}{w_1}^2+12
  {u_1}^3{w_0}^{14}{w_1}^2\nonumber\\&&+8{u_1}^4{w_0}^{13}{w_1}^2+2
  {u_1}^5{w_0}^{12}{w_1}^2+144 r{w_0}^{17}{w_1}+8
  {w_0}^{18}
\end{eqnarray}
Finally for the axial pressure $p_z$ we have
\begin{eqnarray} 
\frac{2 {u_1} {w_1}^2\mathcal{R}}{(w_0 + r w_1)^15}\geq0\\
\mathcal{R}&=& 192 \alpha  {u_1} {w_1}^5 (r {w_1}+{u_1}+{w_0})^2 ({u_1}^2 (r {w_1}+{w_0}) ((7 r+17) {w_1}+7
   {w_0})\nonumber\\&&+{u_1} (r {w_1}+{w_0})^2 ((29 r-71) {w_1}+29 {w_0})+8 (r {w_1}+{w_0})^3 ((3 r-1) {w_1}+3 {w_0})\nonumber\\&&+77 {u_1}^3
   {w_1})+(r {w_1}+{w_0})^{12}
\end{eqnarray}
We plot these inequalities in the Figure 1 (Top-Left panel). As we observe here, the NEC is satisfied for all three pressure components.

\par
$\bullet$ Second case\par
For the second family for $p_r,p_{\varphi},p_z$ we will have respectively
\begin{eqnarray} 
\frac{\mathcal{P}_2}{w_0^8 (r^2 u_0 + w_0)^4}\geq0\\
\mathcal{P}_2&=&1024 \alpha  r^2 {u_0}^5 (56 r^{14} {u_0}^7+360 r^{12} {u_0}^6 {w_0}+920 r^{10} {u_0}^5 {w_0}^2\nonumber\\&&+1128 r^8 {u_0}^4 {w_0}^3-r^6 {u_0}^3
   ({w_0}-552) {w_0}^4-2 r^4 {u_0}^2 {w_0}^5 (3 {w_0}+68)-r^2 {u_0} {w_0}^6 (9 {w_0}+248)\nonumber\\&&-72 {w_0}^7)-4 {u_0} {w_0}^6 (3
   r^2 {u_0}-{w_0}) (r^2 {u_0}+{w_0})^4
\end{eqnarray}

\begin{eqnarray}
\frac{\mathcal{Q}_2}{w_0^8 (r^2 u_0 + w_0)^4}\geq0\\
\mathcal{Q}_2&=&4 w_0^6 (-r^10 u_0^6 - 3 r^8 u_0^5 w_0 - 2 r^5 (3 + r) u_0^4 w_0^2 + \nonumber\\&&
    2 (-3 + r) r^3 u_0^3 w_0^3 + 3 r^2 u_0^2 w_0^4 + (2 + u_0) w_0^5) - 
 1024 r^2 u_0^5 (40 r^14 u_0^7 +\nonumber\\&& 264 r^12 u_0^6 w_0 + 
    760 r^10 u_0^5 w_0^2 + 1272 r^8 u_0^4 w_0^3 + 9 w_0^7 (8 + w_0) + 
    3 r^6 u_0^3 w_0^4 (456 + w_0)\nonumber\\&& + r^4 u_0^2 w_0^5 (952 + 15 w_0) + 
    r^2 u_0 w_0^6 (392 + 25 w_0)) \alpha
\end{eqnarray}

\begin{eqnarray}
\frac{4 u_0 w_0^6 (-r^2 u_0 + w_0) + 
 16384 r^2 u_0^5 (r^2 u_0 + w_0)^2 (2 (-5 + r) r^2 u_0 + 
    3 (-3 + r) w_0) \alpha}{w_0^8}\geq0
\end{eqnarray}
As we observe from panel (Top-Right) the azimuthal pressure violates the NEC.
\par
$\bullet$ Third case\par
For the third family we write

\begin{eqnarray}
&&-\frac{2 e^{2 u_0} (r^{-8 + 2 \beta}) \beta \mathcal{P}_3}{w)^8}\geq0\\
\mathcal{P}_3&=&8 e^{2 u_0} r^{2 \beta} w_0^5 \alpha (3 - 2 \beta)^2 \beta^5 + 
 r^6 w_0^6 (1 + \beta) +  64 e^{6 u_0)} r^(
  6 \beta) \alpha \beta^4 (-3 + 2 \beta) (28 - 33 \beta + 
    6 \beta^2)
\end{eqnarray}

\begin{eqnarray}
&&
-\frac{128 \alpha  \beta ^5 (2 \beta -3) (3 \beta -4) (6 \beta -7) e^{8 {u_0}} r^{8 \beta -8}}{{w_0}^8}-\frac{16 \alpha  \beta ^5 (2 \beta -3) (\beta  (18 \beta
   -47)+28) e^{4 {u_0}} r^{4 \beta -8}}{{w_0}^3}\nonumber\\&&+8 e^{-4 {u_0}} {w_0}^3 r^{-4 \beta }-\frac{2 \beta  e^{2 {u_0}} r^{2 \beta -2}}{{w_0}^2}-\frac{3 \beta
   ^3}{r^3}\geq0
\end{eqnarray}

\begin{eqnarray}
&&
\frac{1024 \alpha  (\beta -1) \beta ^5 (2 \beta -3) e^{8 {u_0}} r^{8 \beta -8} (-6 \beta +r+7)}{{w_0}^8}-\frac{2 \beta  e^{2 {u_0}} r^{2 \beta -2}}{{w_0}^2}\geq0
\end{eqnarray}
Obviously all three satisfy the NEC, as we see in the (Middle-Left).

\par
$\bullet$ Fourth case\par
For the fourth family we derive the following expressions

\begin{eqnarray}
&&-\frac{1120 \sqrt{10} r^{9/2} w_0^6 - 30528 r^3 \alpha + 
  3645 \sqrt{2} w_0^4 \alpha}{3200 r^5 w_0^8}\geq0
\end{eqnarray}

\begin{eqnarray}
&&
\frac{-128 \sqrt{10} r^{9/2} w_0^6 + 90 r^2 w_0^8 (-9 + 160 \sqrt{2} w_0^4) + 
 4032 r^3 \alpha + 1863 \sqrt{2} w_0^4 \alpha}{640 r^5 w_0^8}\geq0
\end{eqnarray}

\begin{eqnarray}
&&
\frac{-5 \sqrt{10} r^{3/2} w_0^6 + 180 \alpha + 72 r \alpha}{25 r^2 w_0^8}\geq0
\end{eqnarray}
As we observe for $r>1$, all the pressure components satisfy the NEC (Middle panel-Right). This special case denotes the existence of a wormhole's like singularity in the cylindrical configurations. The throat's radius is located in a point $r^*\sim 1$.

\par
$\bullet$ Five case\par
For this case we also have:

\begin{eqnarray}
&&\\
&&-64 \alpha  ({u_0}-2)^2 ({u_0}-1)^4 {u_0}^2 {w_1}^8 (r
   {w_1}+{w_0})^{4 {u_0}-11}-2 \left({u_0}^2-1\right) {w_1}^2 (r
   {w_1}+{w_0})^{2 {u_0}-4}\nonumber\\&&-512 \alpha  ({u_0}-2)^2 ({u_0}-1)^2
   {u_0} {w_1}^7 ({w_1} ({u_0} (-2 r+3 ({u_0}-14)
   {u_0}+49)-24)\nonumber\\&&-2 {u_0} {w_0}) (r {w_1}+{w_0})^{8 ({u_0}-2)}\geq0
\end{eqnarray}

\begin{eqnarray}
&&-64 \alpha  ({u_0}-2)^2 ({u_0}-1)^2 {u_0}^2 ({u_0} (9
   {u_0}-32)+21) {w_1}^8 (r {w_1}+{w_0})^{4 {u_0}-11}\nonumber\\&&-512 \alpha 
   ({u_0}-2)^2 ({u_0}-1)^2 {u_0} {w_1}^7 \left({w_1}
   \left({u_0} \left(-2 r+9 {u_0}^2-42 {u_0}+55\right)-24\right)-2
   {u_0} {w_0}\right) (r {w_1}+{w_0})^{8 ({u_0}-2)}\nonumber\\&&+8 (r
   {w_1}+{w_0})^{3-4 {u_0}}+\frac{u_0 (-3 + (5 - 3 u_0) u_0) w_1^3}{(w_0 + r w_1)^3}\geq0
\end{eqnarray}

\begin{eqnarray}
&&
-1024 u_0 (2 - 3 u_0 + u_0^2)^2 w_1^7 (w_0 + r w_1)^(
 8 (-2 + u_0)) (-12 w_1 + 12 u_0^3 w_1 \nonumber\\&&- u_0 (w_0 + (-41 + r) w_1) - 
   2 u_0^2 (w_0 + (20 + r) w_1)) \alpha\geq0
\end{eqnarray}
This case is very interesting. Because it describes a geometry very similar to the wormholes in which the transverse wormhole is formed and the axial component violates the NEC (Bottom-Left).\\
For the model $f(G)=\alpha G^3$,  we observe two forms  which both satisfy the NEC as we observe (Bottom-Right).

\begin{figure*}[thbp]
\begin{tabular}{rl}
\includegraphics[width=7.5cm]{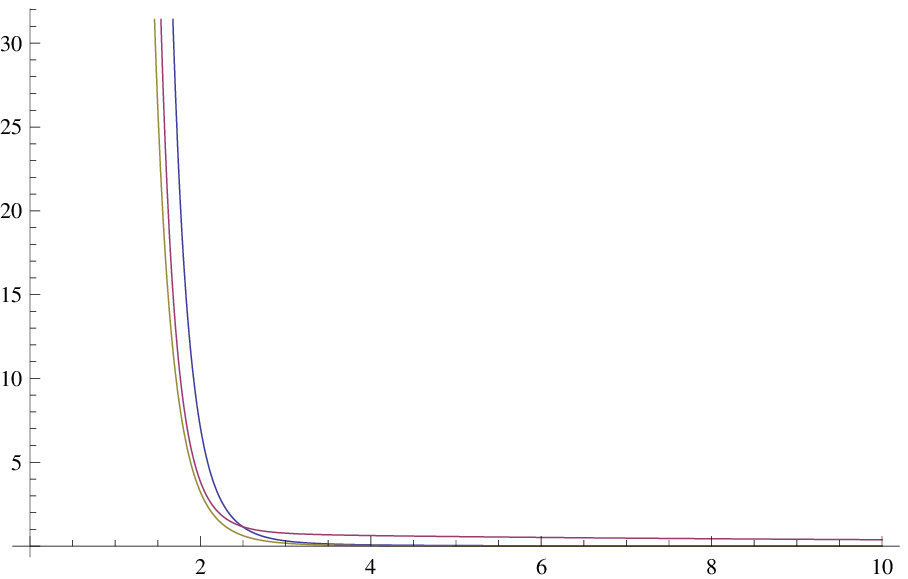}&
\includegraphics[width=7.5cm]{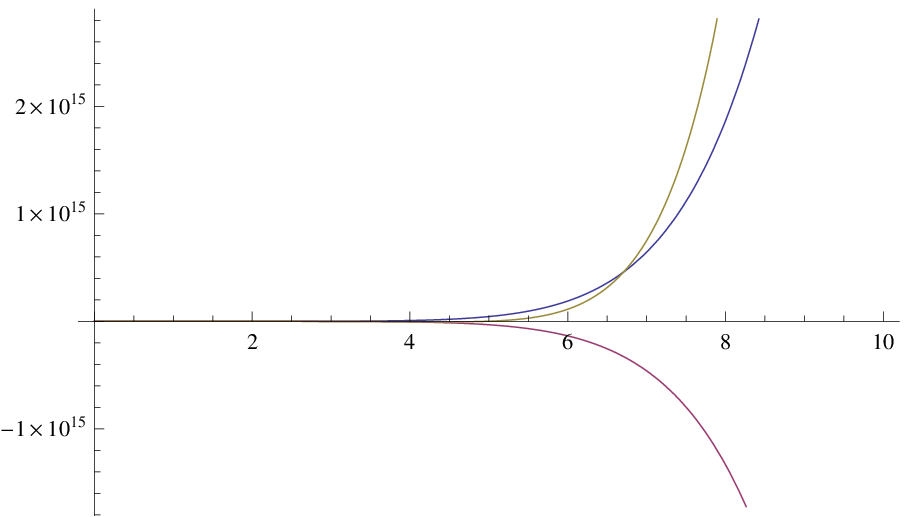} \\
\includegraphics[width=7cm]{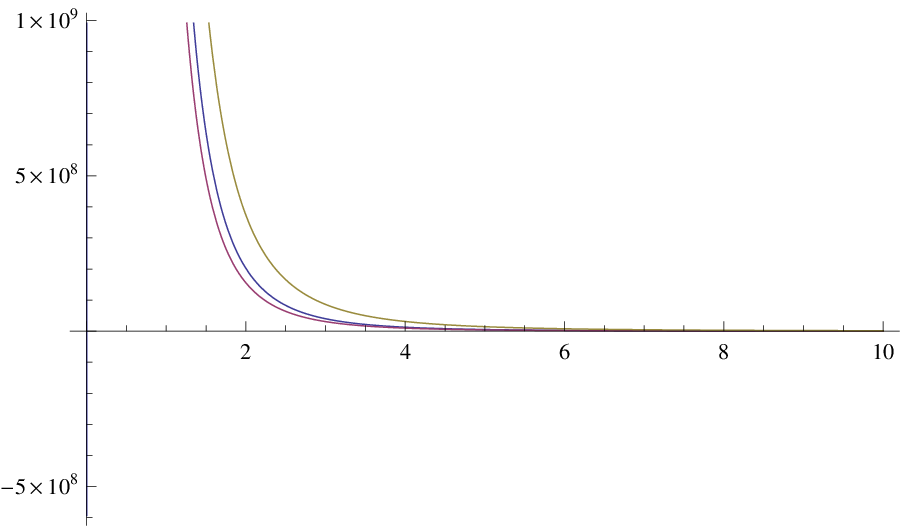}&
\includegraphics[width=7cm]{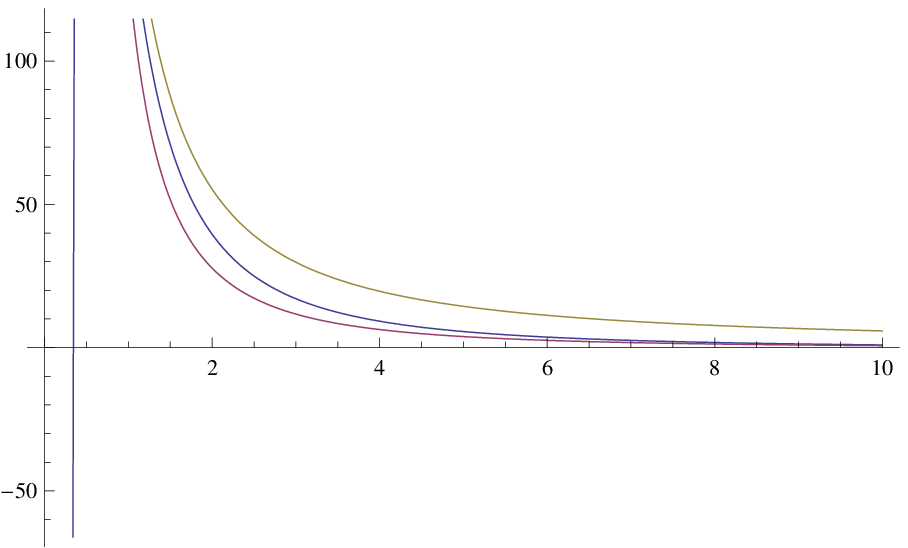} \\
\includegraphics[width=7cm]{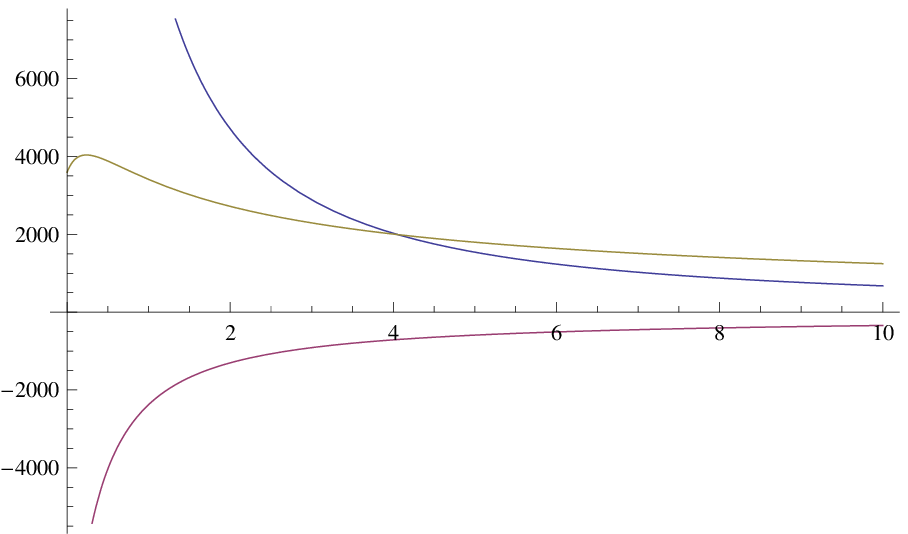}&
\includegraphics[width=7cm]{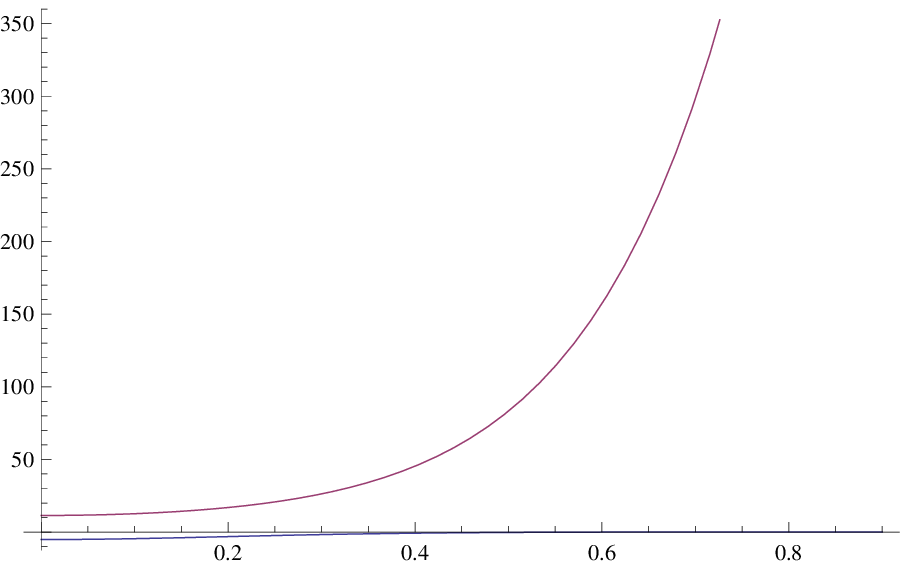} \\
\end{tabular}
\caption{ (\textit{Top Left}) NEC-for model $G^2$-Case1 .   (\textit{Top Right})NEC-for model $G^2$-Case2 . (\textit{Middle Left})NEC-for model $G^2$-Case3 . (\textit{Middle Right})NEC-for model $G^2$-Case4 .  (\textit{Bottom Left})NEC-for model $G^2$-Case5 . (\textit{Bottom Right}) NEC-for model $G^3$. Here the parameters adjusted as $\beta= 0.5, \alpha= 1, u_0 = 1.8, w_0 = 0.7, w_1 = 1.5, u_1 =1.22, k_0 = 1.5$. Blue for radial, red for azimuthal and green for axial.  }
\end{figure*}

\begin{figure*}[thbp]
\includegraphics[width=7.5cm]{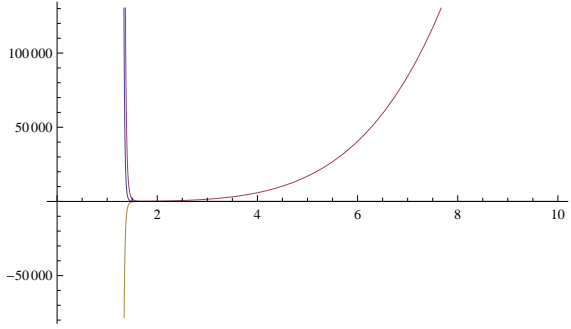}
\caption{NEC for Exponential model; here the parameters adjusted as $\beta= 0.5, \alpha_3= 1, u_0 = 1.8, w_0 = 0.7, w_1 = 1.5, u_1 =1.22, k_0 =u_0^2$. Blue for radial, red for azimuthal and green for axial. }
\end{figure*}

\par
\subsection{Model $f(G)=\alpha_3 \exp{(G)}$}
The last case which is the metric of the exterior of a cosmic string is described by the following expressions:

\begin{eqnarray}
&&2 r^{-12 (1 + k_0)}\Pi_1\geq0\\
\Pi_1&=&-256 {\alpha_3} ({u_0}-1) {u_0} ({k_0}-{u_0})
   ({k_0}-{u_0}+1)^2 \nonumber\\&&\left(-6 {k_0}^2 ({u_0}-1)+{k_0} ({u_0}
   (r+8 {u_0}-13)+6)-2 {u_0} (({u_0}-7) {u_0}+3)\right) r^{4
   {k_0}+8 {u_0}+4} \nonumber\\&&\exp \left(16 ({u_0}-1) {u_0}
   ({k_0}-{u_0}) ({k_0}-{u_0}+1) r^{-4 {k_0}+4
   {u_0}-4}\right)\nonumber\\&&+16384 {\alpha_3} ({u_0}-1)^3 {u_0}^2
   ({k_0}-{u_0})^3 ({k_0}-{u_0}+1)^4 \nonumber\\&&r^{12 {u_0}} \exp \left(16
   ({u_0}-1) {u_0} ({k_0}-{u_0}) ({k_0}-{u_0}+1) r^{-4
   {k_0}+4 {u_0}-4}\right)+{\alpha_3} ({u_0}-1) {u_0} r^{12
   {k_0}+9} ({k_0}-{u_0}) \nonumber\\&&({k_0}-{u_0}+1) \left(-\exp \left(16
   ({u_0}-1) {u_0} ({k_0}-{u_0}) ({k_0}-{u_0}+1) r^{-4
   {k_0}+4 {u_0}-4}\right)\right)\nonumber\\&&+8 {\alpha_3} ({u_0}-1) {u_0}
   ({k_0}-{u_0}) ({k_0}-{u_0}+1) r^{8 {k_0}+4 {u_0}+8} \exp
   \left(16 ({u_0}-1) {u_0} ({k_0}-{u_0})({k_0}-{u_0}+1)
   r^{-4 {k_0}+4 {u_0}-4}\right)+\nonumber\\&& \left({k_0}-{u_0}^2\right) r^{2 (5
   {k_0}+{u_0}+5)}
\end{eqnarray}

\begin{eqnarray}
&&r^{-4 (3 + 3 k_0 + u_0)}\Pi_2\geq0\\
\Pi_2&=&-{u_0} r^{12 {k_0}+4 {u_0}+9} (2 {\alpha_3} ({u_0}-1)
   ({k_0}-{u_0}) ({k_0}-{u_0}+1) \exp (16 ({u_0}-1)
   {u_0} ({k_0}-{u_0}) ({k_0}-{u_0}+1) r^{-4 {k_0}+4
   {u_0}-4}))\nonumber\\&&+(2 {k_0}^2-5 {k_0} {u_0}+{k_0}+3
   {u_0}^2)-512 {\alpha_3} ({u_0}-1) {u_0}
   ({k_0}-{u_0})\nonumber\\&& ({k_0}-{u_0}+1)^2 (({k_0}+1) {u_0}^2
   (8 {k_0}^2 ({k_0}+1)+11 r^3)-{u_0}^3 (8 {k_0}
   ({k_0}+1) ({k_0} ({k_0}+5)+2)+5 r^3))\nonumber\\&&+(r^3 {u_0} ({k_0}
   (-6 {k_0}+r-16)-6))\nonumber\\&&(+6 {k_0} ({k_0}+1) r^3+8 (4 {k_0}+3)
   {u_0}^6-8 (2 {k_0} (3 {k_0}+5)+3) {u_0}^5+8 (4 {k_0}
   ({k_0}+1) ({k_0}+2)+1) {u_0}^4-8 {u_0}^7)\nonumber\\&& r^{4 {k_0}+12
   {u_0}+1} \exp (16 ({u_0}-1) {u_0} ({k_0}-{u_0})
   ({k_0}-{u_0}+1) r^{-4 {k_0}+4 {u_0}-4}))\nonumber\\&&(+16 {\alpha_3}
   ({u_0}-1) {u_0} ({k_0}-{u_0}) ({k_0}-{u_0}+1) (4
   {u_0} ({k_0}-{u_0}+1) (4 {k_0}^2+{k_0} (6-8
   {u_0})+{u_0} (4 {u_0}-7))+r^3))\nonumber\\&&( r^{8 {k_0}+8
   {u_0}+5} \exp (16 ({u_0}-1) {u_0} ({k_0}-{u_0})
   ({k_0}-{u_0}+1) r^{-4 {k_0}+4 {u_0}-4})+32768 {\alpha
   _3} ({u_0}-1)^3 {u_0}^2)\nonumber\\&&( ({k_0}-{u_0})^3 ({k_0}-{u_0}+1)^4
   r^{16 {u_0}} \exp (16 ({u_0}-1) {u_0} ({k_0}-{u_0})
   ({k_0}-{u_0}+1) r^{-4 {k_0}+4 {u_0}-4})+8 r^{16
   {k_0}+11}
\end{eqnarray}

\begin{eqnarray}
&&512{\alpha_3} ({u_0}-1){u_0} ({k_0}-{u_0}) ({k_0}-{u_0}+1)^2 r^{8{u_0}-12 ({k_0}+1)}\nonumber\\&& (64 ({u_0}-1)^2{u_0} ({k_0}-2
  {u_0}) ({k_0}-{u_0}) ({k_0}-{u_0}+1)^2 r^{4{u_0}})\nonumber\\&&-(r^{4{k_0}+4}(-6{k_0}^2 ({u_0}-1)+{k_0} ({u_0} (r+18{u_0}-23)+6)+2
  {u_0} ({u_0} (r-6{u_0}+11)-6)))\nonumber\\&& \exp (16 ({u_0}-1){u_0} ({k_0}-{u_0}) ({k_0}-{u_0}+1) r^{-4{k_0}+4
  {u_0}-4})\geq0
\end{eqnarray}
From Figure.2, we observe that for $r>2$, the energy conditions of type NEC is satisfied.
This analysis completed the dynamical features of the solutions.

\section{Conclusion}
Cylindrical solutions have a key role in the string theory in relation to the spontaneously symmetry breaking of the abelian gauge fields and also as the most fundamental objects which had been produced in the very early universe with a very small mass order. Different kinds of the cosmic strings have been obtained before in Einstein gravity, extended modified gravities both in the classical and also in quantum levels.
In this paper we investigated cylindrically symmetric solutions of a type of modified Gauss-Bonnet gravity. By the choice of some viable forms of the Gauss-Bonnet term, we solved the equations and found seven classes of solutions, including the metric of the exterior of a cosmic string. Further, for checking the dynamical behaviours of the system, we checked the null energy conditions which implies another types of the energy conditions. In some models, the azimuthal pressure violated the energy condition for a distinct radius scale. We interpreted this radius as a candidate for the existence of the cylindrical wormholes in Einstein-Gauss-Bonnet gravity. We mention that in this paper just some particular solutions have been found and the full description of the system, under perturbations, attractors and etc needs more future works.

\vspace{0.5cm}
{\bf Acknowledgement:}  M. J. S. Houndjo  thanks CNPq/FAPES for financial support and M. E. Rodrigues thanks a lot UFES for the hospitality during the elaboration of this work.

\end{document}